\documentclass{aastex62}
\usepackage{amsmath,bm}
\graphicspath{{./}{figures/}}

\submitjournal{AJ}

\shorttitle{Autoencoding SDSS Spectra}
\shortauthors{Portillo et al.}
\turnoffedit
\begin{document}

\title{Dimensionality Reduction of SDSS Spectra with Variational Autoencoders}

%% you desire. In v6.2 you can include a footnote in the title.

\author[0000-0001-8132-8056]{Stephen K. N. Portillo}
\affiliation{DIRAC Institute, Department of Astronomy\\
University of Washington\\
3910 15th Ave NE\\
Seattle, WA 98195, USA}

\author{John K. Parejko}
\affiliation{DIRAC Institute, Department of Astronomy\\
University of Washington\\
3910 15th Ave NE\\
Seattle, WA 98195, USA}

\author[0000-0001-6699-4181]{Jorge R. Vergara}
\affiliation{Department of Computing, Universidad Tecnol\'{o}gica Metropolitana, 
Santiago, Chile}
\affiliation{Millennium Institute of Astrophysics, Santiago, Chile}

\author[0000-0001-5576-8189]{Andrew J. Connolly}
\affiliation{DIRAC Institute, Department of Astronomy\\
University of Washington\\
3910 15th Ave NE\\
Seattle, WA 98195, USA}

\correspondingauthor{Stephen K. N. Portillo}
\email{sportill@uw.edu}

\begin{abstract}

High resolution galaxy spectra contain much information about galactic physics, but the high dimensionality of these spectra makes it difficult to fully utilize the information they contain. We apply variational autoencoders (VAEs), a non-linear dimensionality \edit1{reduction} technique, to a sample of spectra from the Sloan Digital Sky Survey. In contrast to Principal Component Analysis (PCA), a widely used technique, VAEs can capture non-linear relationships between latent parameters and the data. We find that a VAE can reconstruct the SDSS spectra well with only six latent parameters, outperforming PCA with the same number of components. Different galaxy classes are naturally separated in this latent space, without class labels having been given to the VAE. The VAE latent space is interpretable because the VAE can be used to make synthetic spectra at any point in latent space. For example, making synthetic spectra along tracks in latent space yields sequences of realistic spectra that interpolate between two different types of galaxies. Using the latent space to find outliers may yield interesting spectra: in our small sample, we immediately find unusual data artifacts and stars misclassified as galaxies. In this exploratory work, we show that VAEs create compact, interpretable latent spaces that capture non-linear features of the data\edit1{. While a VAE takes substantial time to train ($\approx$1 day for 48000 spectra), once trained,}  VAEs can enable the fast exploration of large astronomical data sets.
\end{abstract}

%% As of Monday, June 3rd, 2019, the AAS journals will categorize
%% articles using Unified Astronomy Thesaurus (UAT) concepts instead
%% of the old subject keywords system
%% \keywords{galaxies: fundamental parameters –-- methods: data analysis –-- techniques: spectroscopic}

\section{Introduction}
\label{sec:intro}

The galaxy spectra in the Sloan Digital Sky Survey (SDSS) contain a wealth of information about the physical processes occurring in galaxies, but to fully make use of these spectra, methods are needed that can handle both the complexity of galaxy spectra and the vast number of available spectra. Spectra consist of thousands of flux measurements that depend on galaxy properties in complex ways. To investigate physical trends in galaxies, scientists must extract statistics from each spectrum that are easily related to the physical properties of the galaxy. In other words, scientists reduce the dimensionality of the data in a way that enables them to probe the galaxies' properties.

Spectra can be fit with theoretical or semi-analytic models, yielding physically interpretable parameters for each spectrum. This approach has the advantage of directly connecting the observations to a physical model. Incomplete models can, however, introduce biases to the inferred parameters or miss interesting behavior not included in the model. 

To supplement model-based interpretation of the data, empirical methods are desirable. One approach is to focus on parts of the spectrum that are known to be physically significant, like emission lines. Line ratio tests \citep{baldwin_classification_1981,osterbrock_optical_1985,kewley_theoretical_2001,kewley_host_2006,kewley_understanding_2019} leverage the sensitivity of emission lines to ionization levels while minimizing the effect of foreground dust extinction. These tests are very useful in classifying star-forming galaxies and active galactic nuclei, but ignore the information present in the continuum of the spectrum.

Another approach is to use dimensionality reduction or classification techniques that extract useful features from the data without any prior astrophysical knowledge. These methods include Principal Component Analysis (PCA), Independent Component Analysis \citep{lu_ensemble_2006}, Local Linear Embedding \citep{vanderplas_reducing_2009}, and neural net classification \citep{folkes_artificial_1996,ball_galaxy_2004}. Of these methods, PCA is one of the most successful and widely used. PCA finds components that can be added in linear combinations to best approximate the input data. The coefficients of the linear combination that approximates a given spectrum can be thought of as a compression of the spectrum.  \cite{yip_distributions_2004} found that the coefficients of the first three PCA components of the SDSS spectra are sufficient to separate early-type galaxies, late-type galaxies, and extreme emission-line galaxies. They also found that the first eight PCA components are sufficient to reconstruct the continuum levels and spectral line ratios of all but the most extreme emission-line galaxies, representing a considerable reduction in dimensionality from the 3839 pixels in each spectrum. Because the PCA components are simply added together to yield spectra, each component can be interpreted. For example, the first PCA component removes blue continuum and nebular emission lines and adds continuum light, suggesting that this component is negatively correlated with star formation activity.

PCA's linearity, however, inhibits its ability to efficiently reduce the dimensionality of data when non-linear features are present. \cite{yip_spectral_2004} apply PCA to SDSS quasars out to $z=5.41$ and find that 50 components are necessary to acceptably reconstruct a typical spectrum. By binning the quasars by redshift and luminosity and performing PCA separately in each bin, they can get a similar reconstruction with only 10 components. They also find that broad absorption lines cannot be captured by a single PCA component, but instead are reconstructed as the combination of many components. %10.1086/497973 fits broad-line AGN spectra using a combination of PCA components describing galaxies from \cite{yip_distributions_2004} and PCA components from \cite{yip_spectral_2004} for low-redshift quasars.

There are many empirical dimensionality reduction and classification techniques that are non-linear, giving them more flexiblity than PCA in dealing with non-linear features. These techniques include diffusion maps \citep{richards_exploiting_2009}, local linear embedding \citep{vanderplas_reducing_2009}, k-means clustering \citep{almeida_automatic_2010}, locally-biased semi-supervised eigenvectors \citep{lawlor_mapping_2016}, self organizing maps \citep{meusinger_unusual_2012,meusinger_large_2017}, \edit1{and random forests \citep{2017MNRAS.465.4530B,2018MNRAS.476.2117R}}.

Autoencoders are a class of neural network that has been widely studied in the machine learning literature and that astronomers have started to adopt. For example, autoencoders have been used to: estimate stellar atmospheric properties from spectra \citep{yang_autoencoder_2015,li_parameterizing_2017}, identify spatial structures in Tycho's supernova remnant using X-ray spectra \citep{iwasaki_x-ray_2019}, morphologically classify radio AGN \citep{ma_machine_2019}, classify variable star light curves \citep{naul_recurrent_2018,tsang_deep_2019}, and emulate cosmological simulations \citep{chardin_deep_2019,troster_painting_2019}.

The goal of this work is to use variational autoencoders (a specific subclass of autoencoders) to address the limitations of PCA. A key feature of PCA is that it focuses on reconstruction: the PCA coefficients for an input are a low-dimension representation that can be used to reconstruct an approximation to that input. We can then compare the reconstruction to the input to see what features the PCA components are capturing. Not only can we reconstruct observed spectra: any combination of PCA coefficients can be used to construct new, synthetic spectra. In the case of PCA, any synthetic spectra will simply be linear combinations of the PCA components. Similarly, VAEs have an encoder that returns a low-dimension latent representation of a given input and a decoder that returns a reconstruction given that latent representation. We can use the decoder to construct a synthetic spectrum for any point in the latent space. Unlike PCA, the VAE's non-linearity allows it to find non-linear relationships between the latent representations and the spectra. We introduce autoencoders and variational autoencoders in \autoref{sec:vaes}. Then in \autoref{sec:application} we apply variational autoencoders to a subset of the SDSS spectra and consider the quality of the VAE reconstructions and the intepretability of the latent representations. We discuss the advantages and challenges of using VAEs as a dimensionality reduction technique in \autoref{sec:discussion} before concluding in \autoref{sec:conclusion}.

\section{Variational Autoencoders}
\label{sec:vaes}

\begin{figure}
    \plotone{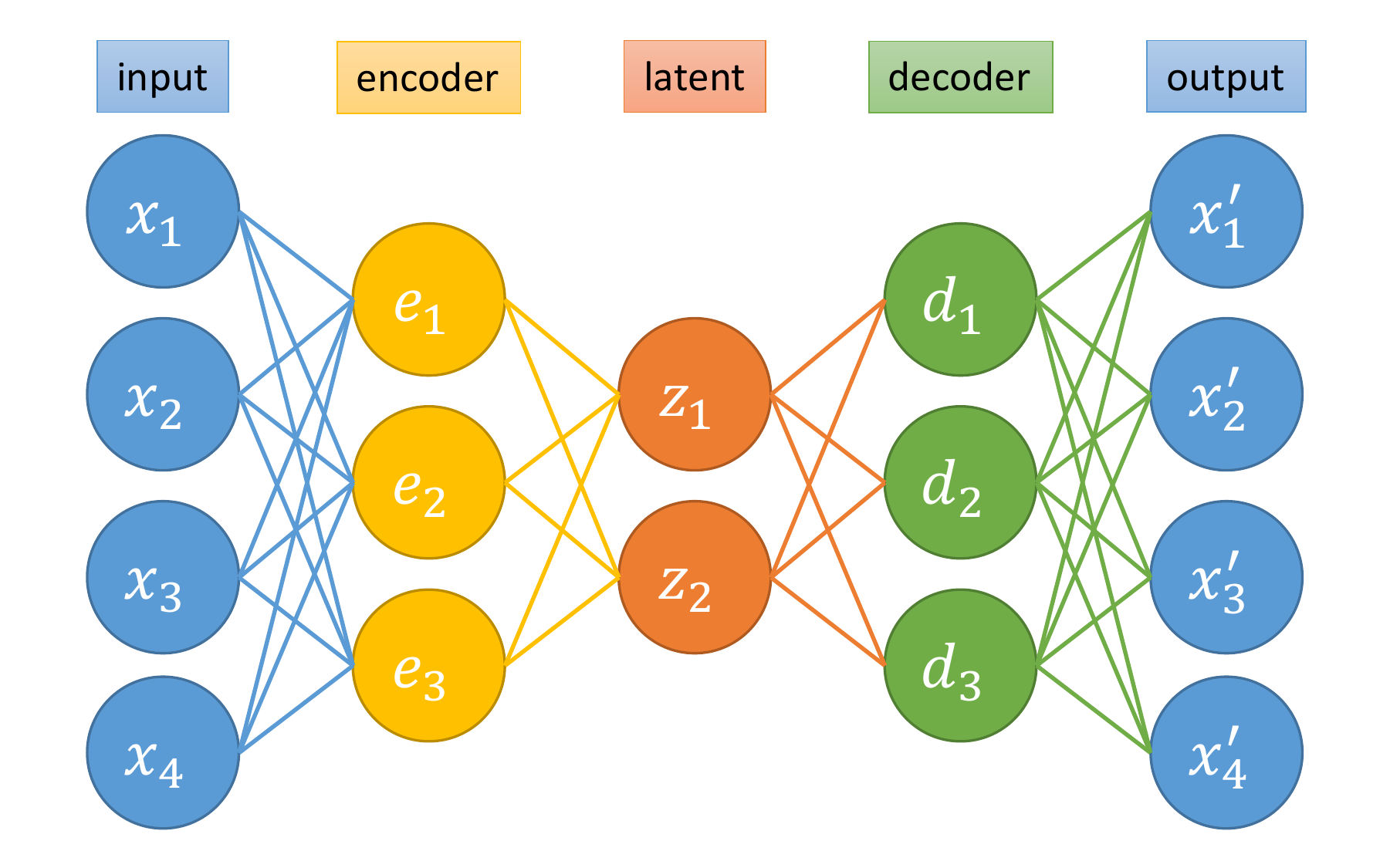}
    \caption{Structure of an autoencoder with a 4-3-2-3-4 architecture. It compresses four-dimensional inputs to a two-dimensional latent space and both the encoder and decoder are single layers with three neurons.}
    \label{fig:AEdiagram}
\end{figure}

Autoencoders are feed-forward neural networks that learn efficient encodings of data in an unsupervised manner. An autoencoder consists of two parts: an encoder that takes data as input and compresses it to produce a \edit1{latent representation}, followed by a decoder that takes the latent representation and decompresses it to produce a reconstruction of the original data. \edit1{In an undercomplete\footnote{There are autoencoder architectures that do not require the latent space to be lower-dimensional than the data, but we focus on undercomplete autoencoders in this work.} autoencoder,} the latent representation is lower-dimensional than the data, \edit1{so} the autoencoder cannot simply learn the identity function. \edit1{An autoencoder that learns the identity function would not be useful: its reconstructions would simply be a blind copy of the input and it would not have learned a compact representation of the data.} Instead, it must find a compression that allows the input data to be approximately reconstructed. This compression is tailored for the data set that the autoencoder is trained on, thus the latent representation often reflects meaningful properties of the data. The latent representations can be thought of as occupying a latent space: this space is the range of the encoder and domain of the decoder. PCA reconstruction can be thought of as a restricted autoencoder. In this case, the encoder projects the data onto the first $n$ PCA components, giving a low-dimensional latent representation: the projection coefficients. The corresponding decoder then returns the linear combination of the first $n$ PCA components with these coefficients as the reconstruction of the input. Training PCA reconstruction consists of finding the first $n$ PCA components: of all the possible bases of $n$ vectors, the first $n$ PCA components are the basis that minimizes the mean squared reconstruction error. In an autoencoder, the encoder and decoder can learn more complex operations than projection and linear combination, allowing the autoencoder to capture non-linear features of the data. We will outline how autoencoders work in this section, but more details can be found in machine learning references such as \cite{Goodfellow-et-al-2016} (Chapter 14).

A feed-forward neural network consists of neurons arranged in a sequence of layers, where the activation of a neuron depends only on the activations of the neurons in the previous layers. \autoref{fig:AEdiagram} depicts a simple autoencoder that takes in a four-dimensional input $\mathbf{x}$, compresses it to a two-dimensional latent representation $\mathbf{z}$, and then outputs a four-dimensional reconstruction $\mathbf{x'}$. \edit1{In our case, this autoencoder would take in spectra with four pixels and compress each to a two-dimensional representation before reconstructing the spectrum's four pixel values.} This autoencoder has five layers (arranged from left to right in \autoref{fig:AEdiagram}) with 4, 3, 2, 3, and 4 neurons, respectively, so it has a 4-3-2-3-4 architecture. The first layer is the input layer: when one of the four-dimensional input data is fed into the autoencoder, the four input layer neurons' activations are set to the four values in the input data. The encoder consists of a single fully connected layer, meaning that each of the activations of its three neurons depends on the activations of all of the input neurons. Specifically, writing the input layer activations as a vector $\mathbf{x}$, the vector of encoder neuron activations $\mathbf{e}$ is:
\begin{equation}
    \mathbf{e} = f(W^{(e)} \mathbf{x} + \mathbf{b}^{(e)})
\end{equation}
where $W^{(e)}$ is a matrix of weights, $\mathbf{b}^{(e)}$ is a vector of biases, and $f$ is the non-linear activation function that is applied element-wise to its argument. A common choice is the rectified linear unit (ReLU)
\begin{equation}
    \mathrm{ReLU}(y) = 
    \begin{cases}
        y & y \ge 0 \\
        0 & y < 0
    \end{cases}
\end{equation}
because it is simple to take the derivative of, which is useful when training the network. While the activation function for a single neuron may be simple, having many neurons in a layer and many layers in the network gives the neural network considerable expressive power. \edit1{The encoder neuron activations for a given input can be interpreted as a partially compressed representation of that input.} The next layer, the latent layer, is fully-connected to the encoder layer, with the latent neuron activations being given by:
\begin{equation}
    \mathbf{z} = W^{(z)} \mathbf{e} + \mathbf{b}^{(z)}
\end{equation}
where $W^{(z)}$ is a matrix of weights, $\mathbf{b}^{(z)}$ is a vector of biases, and there is no activation function. \edit1{We note that the weights and biases can be different for each layer.} \edit1{The latent neuron activations for a given input comprise the autoencoder's most compact representation of that input. The latent neuron activations are analogous to the PCA coefficients, which are the compact representation that PCA produces.}

The decoder is also a single layer of three neurons, fully connected to the latent layer, with their activations being given by
\begin{equation}
    \mathbf{d} = f(W^{(d)} \mathbf{z} + \mathbf{b}^{(d)})
\end{equation}
where $W^{(d)}$ is a matrix of weights, $\mathbf{b}^{(d)}$ is a vector of biases, and $f$ is the activation function. \edit1{The decoder neuron activations can be interpreted as a partially decompressed version of the latent neuron activations.} Finally, the output reconstruction is given by the activation of the last layer:
\begin{equation}
    \mathbf{x'} = W^{(x')} \mathbf{d} + \mathbf{b^{(x')}}
\end{equation}
where $W^{(x')}$ is a matrix of weights, $\mathbf{b}^{(x')}$ is a vector of biases, and there is no activation function. \edit1{The output neuron activations are the autoencoder's reconstruction of the input data. Each neuron in this layer maps one-to-one with a neuron in the input layer, and thus with one of the dimensions of the data (one of the spectrum pixels, in our case).}

The autoencoder is trained to minimize \edit1{a defined} reconstruction loss \edit1{which decreases when the output is closer to the input, like the $\chi^2$ test statistic between the output and input}. That is, the weights $W^{(e,z,d,x')}$ and biases $\mathbf{b}^{(e,z,d,x')}$ are iteratively changed to decrease the average reconstruction loss on some training data set. Finding the optimal weights and biases is a high-dimensional and non-linear optimization problem that requires specialized optimizers. Optimizers often make use of the gradient of the reconstruction loss with respect to the weights and biases. Simple activation functions like ReLU make these gradients fast to calculate. Often, the gradient of the average loss is not calculated on the entire training data set, but on random subsets of the training set called mini-batches. This subsampling adds some stochasticisty to the gradient, which can help the optimizer avoid local minima. Also, optimizers often allow step sizes to be adaptively changed for each weight and bias during training. One popular optimizer, Adam \citep{kingma_adam:_2015}, keeps a moving average of the gradient and gradient squared to determine good step sizes as training progresses. In finding a combination of weights and biases that minimize the reconstruction error, the autoencoder is implicitly learning a dimensionality reduction of the data. These weights and biases define the encoder that maps input data to a lower-dimensional latent representation as well as the decoder that produces a reconstruction from a latent representation.

\begin{figure}
    \plotone{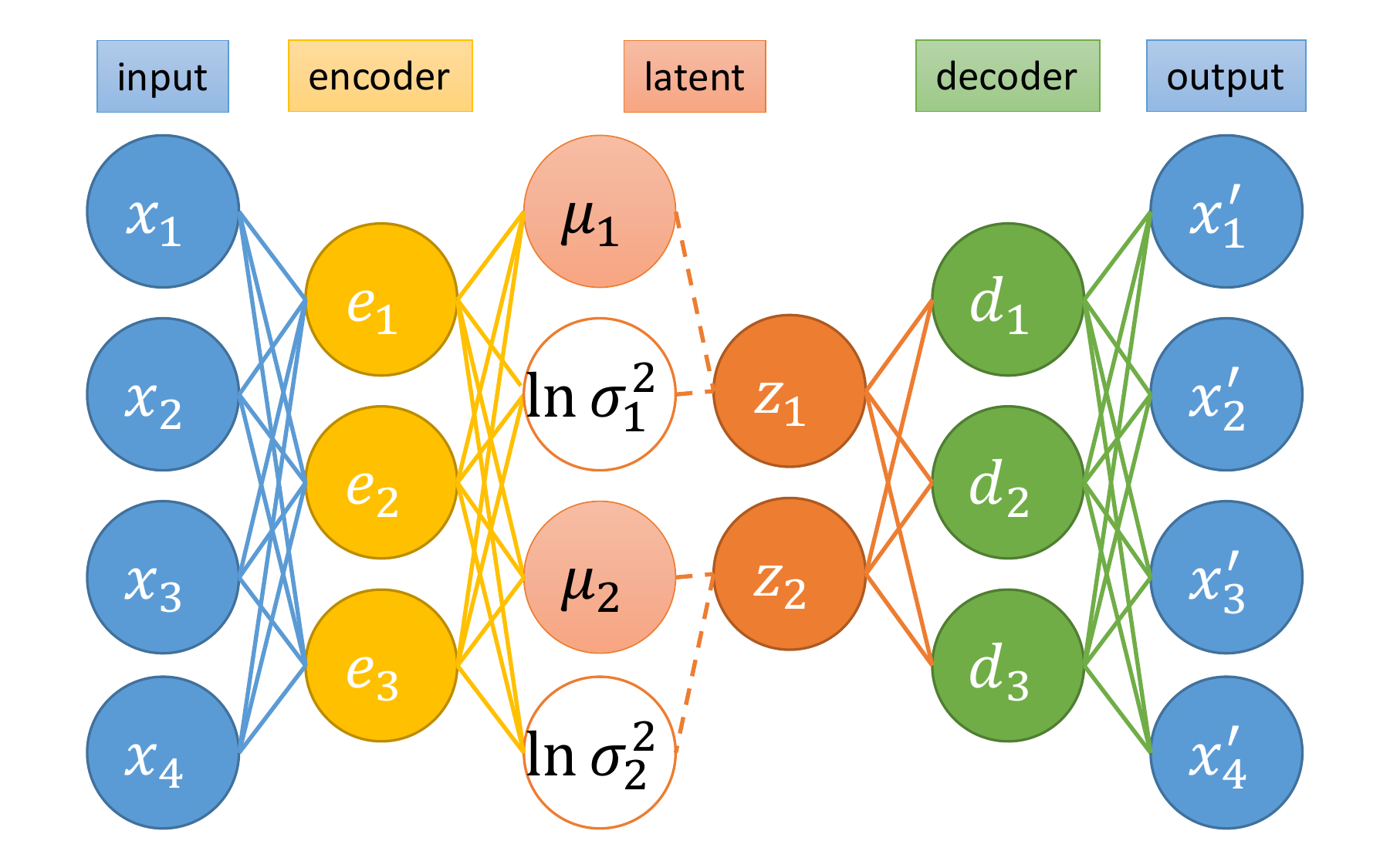}
    \caption{Structure of a variational autoencoder with a 4-3-2-3-4 architecture (similar structure to the autoencoder in \autoref{fig:AEdiagram}). The dashed lines indicate that each latent value $z_{1,2}$ is drawn from a Gaussian distribution defined by a latent mean $\mu_{1,2}$ and log variance $\ln \sigma_{1,2}^2$.}
    \label{fig:VAEdiagram}
\end{figure}

In a variational autoencoder \citep{kingma_auto-encoding_2013}, inputs get mapped onto a distribution in latent space, rather than a single point. Without this change, the latent space may not be continuous, as inputs can simply be mapped to disjoint points in latent space. By mapping inputs to distributions, similar inputs are allowed to have overlapping distributions, allowing interpolation between them. \cite{kingma_auto-encoding_2013} map each input onto a multivariate Gaussian distribution with no correlations. The latent layer is replaced with two sets of neurons: one representing the means in each of the dimensions of latent space, the other representing the log variances (log variances are used to guarantee that the variances are positive). We will refer to these values as the latent means and latent log variances. The decoder samples a point in latent space from this Gaussian distribution and then decodes it as the reconstruction. \autoref{fig:VAEdiagram} shows a VAE with a similar architecture to the autoencoder in \autoref{fig:AEdiagram}. The latent means and latent variances each have their own weights and biases
\begin{eqnarray}
    \bm{\mu} &=& W^{(\mu)} \mathbf{e} + \mathbf{b}^{(\mu)} \\
    \ln \bm{\sigma^2} &=& W^{(\ln \sigma^2)} \mathbf{e} + \mathbf{b}^{(\ln \sigma^2)}
\end{eqnarray}
and a point in latent space is sampled \edit1{from $q(z|x)$, the Gaussian with these latent means and variances}
\begin{equation}
    \mathbf{z} \sim q(\mathbf{z}|\mathbf{x}) \equiv \mathcal{N}(\bm{\mu}, \mathrm{diag}(\exp(\ln \bm{\sigma^2}))).
\end{equation}
\edit1{This Gaussian can be thought of as a distribution of latent representations that are consistent with the input. Instead of propagating this distribution forward, the decoder samples a point from it and propagates it:}
\begin{equation}
    \mathbf{d} = f(W^{(d)} \mathbf{z} + \mathbf{b}^{(d)}).
\end{equation}
Note that this sampling step is not deterministic, but as the VAE is trained, the same input will be seen many times, and many samples in latent space will be propagated through the decoder. A prior is placed on the latent space that the latent distributions, averaged over the entire data set, comprise a standard multivariate Gaussian. \edit1{The latent distributions for each input are already Gaussians, but this prior acts to regularize the means and variances of each latent distribution so that they roughly add up to a standard multivariate Gaussian for the entire data set. Different priors would give different regularizations, but a standard multivariate Gaussian is commonly used to make training the VAE faster.} The objective function used is the evidence lower bound (ELBO), which is the sum of the reconstruction loss and the \edit1{Kullback--Leibler (KL) divergence \citep{kullback1951}} between the latent distribution for the input $q(\mathbf{z}|\mathbf{x})$ and the prior $p(\mathbf{z})$:
\begin{equation}
    \mathrm{ELBO} = L(\mathbf{x}, \mathbf{x'}) + D_{\mathrm {KL}}(q(\mathbf{z}|\mathbf{x})||p(\mathbf{z})).
\end{equation}
The KL divergence between two probability distributions $p$ and $q$ is a measure of the difference between the two distributions and is defined as
\begin{equation}
D_{\mathrm{KL}}(q||p) = \int q(z) \log\left(\frac{q(z)}{p(z)}\right) dz.
\end{equation}
The reconstruction loss can be any function measuring the difference between the input and the reconstruction, like a $\chi^2$ loss. If the reconstruction loss is a negative log likelihood (like the $\chi^2$ loss for inputs with Gaussian errors) and the decoder is thought of as a generative model for the data, then the encoder is performing variational inference. That is, the Gaussian described by the encoder is the closest (by KL divergence) such Gaussian to the posterior that is implied by the combination of the prior, decoder (generative model), and reconstruction loss (negative log likelihood).

We use InfoVAE \citep{Zhao2019InfoVAEBL}, a VAE variant that addresses two issues with the ELBO objective function. First, the KL divergence term is not strong enough to discourage the VAE from mapping different inputs to disjoint distributions. This problem is worst when the dimensionality of the input is much greater than the dimensionality of the latent space. Second, the KL divergence term is minimized when the latent distribution for each input matches the prior. In this case, the latent distribution does not depend on the input at all, as it always matches the prior. If this term is strengthened by multiplying it by a large number, as in $\beta$-VAE \citep{higgins_-vae:_2017}, then this behavior will cause the VAE to under-utilize the latent space as it is encouraged to match the latent distribution to the prior. InfoVAE uses an objective function that alleviates these problems: 
\begin{equation}
    L_{\mathrm{InfoVAE}} = L(\mathbf{x}, \mathbf{x'}) + (1-\alpha) D_{\mathrm{KL}} (q(\mathbf{z}|\mathbf{x})||p(\mathbf{z})) + (\alpha + \lambda - 1) D_{\mathrm{MMD}} (q(\mathbf{z})||p(\mathbf{z})).
\end{equation}
With $\alpha = 0$, the first two terms are the same as the ELBO objective. The last term compares $q(\mathbf{z})$, the latent distribution averaged over inputs, with the prior. The latent distribution averaged over inputs is a sum of Gaussians, but its KL divergence with the prior would be expensive to calculate. Instead, samples from the latent distributions of a minibatch of inputs are drawn, alongside samples from the prior. The \edit1{Maximum Mean Discrepancy \citep[MMD;][]{NIPS2006_3110}} is then calculated between the samples from the latent distributions and the prior, given by
\begin{equation}
    D_{\mathrm{MMD}} = \frac{1}{m^2}\sum_{i,j=1}^{m}\kappa(u_i,u_j) - \frac{2}{mn}\sum_{i,j=1}^{m,n}\kappa(u_i,v_j) + \frac{1}{n^2}\sum_{j=1}^{n}\kappa(v_i,v_j)
\end{equation}
where $u_{1...m}$ are samples from the latent distributions, $v_{1...n}$ are samples from the prior, and $\kappa$ is a positive-definite kernel function such as a squared exponential. The MMD term encourages the latent distribution averaged over all inputs (rather than just the distribution for a single input) to match the prior, and can be strengthened by choosing $\lambda > 1$.

\section{Application}
\label{sec:application}

As an initial demonstration of VAEs, we train a VAE using a subset of $\approx$ 64000 spectra taken from the SDSS DR7 \citep{york_sloan_2000,strauss_spectroscopic_2002,gunn_2.5_2006,abazajian_seventh_2009,smee_multi-object_2013}. Following the approach of \cite{vanderplas_reducing_2009}, we select spectra from the main galaxy ($\approx$ 59000 spectra) and quasar ($\approx$ 4500 spectra) samples with redshifts $z<0.36$ using the AstroML \citep{vanderplas_introduction_2012} SDSS query generator\footnote{The query we used was \texttt{SELECT TOP 64000 plate, mjd, fiberid FROM specObj WHERE ((PrimTarget \& (TARGET\_GALAXY + TARGET\_QSO\_CAP + TARGET\_QSO\_SKIRT)) > 0) AND (z > 0) AND (z <= 0.36)}}. We then use AstroML to pre-process the selected spectra. First, we shift all spectra to their rest frames, resample them to 1000 logarithmically spaced wavelength pixels between 3388 {\AA} - 8318 {\AA}, and normalize by total flux in this wavelength range. This resampling is much coarser than the SDSS spectral resolution, and so small-scale information is lost. Since this work is meant to be a first demonstration of VAEs on SDSS spectra, we accept this loss of information in order to reduce the computational cost of training our VAEs. Then, we infill bad pixels (due to e.g. bad sky line subtraction) using an iterative PCA procedure, as done in \cite{yip_distributions_2004}. We then only keep spectra that are classified by SDSS's spectroscopic pipeline \texttt{spectro1d} \citep[subsection 4.10]{stoughton_sloan_2002} as galaxies (\texttt{SPEC\_CLN = 2}) or quasars (\texttt{SPEC\_CLN = 3}), excluding 2\% of the spectra. A validation set of $\approx$ 16000 spectra of the remaining spectra is set aside, leaving $\approx$ 47000 spectra in the training set.

For the reconstruction loss, we use a Gaussian log likelihood using the reported uncertainties for each spectrum. Bad pixels have their weight in the loss is set to zero (which is equivalent to giving them infinite uncertainty). To prevent a small number of very high SNR pixels from dominating the total loss, an uncertainty floor is added to each good pixel that caps the maximum pixel SNR to 50. This cap affects 1\% of pixels across all spectra and only 0.5\% of spectra have a median pixel SNR greater than 50.

We implement InfoVAE using \texttt{pytorch} \citep{paszke_automatic_2017} and train our VAEs using \texttt{pytorch}'s built-in Adam optimizer \citep{kingma_adam:_2015}, with a batch size of 64 and a starting learning rate of $10^{-3}$. When the objective function does not improve for five epochs on the validation set, we reduce the learning rate by a factor of 10. We stop training  if the objective function does not improve for ten epochs on the validation set (ie. early stopping). For the InfoVAE objective function hyperparameters, we set $\alpha=0$ (as recommended by \cite{Zhao2019InfoVAEBL} for a simple decoding distribution) and use random search for $\lambda$. We use two hidden layers each for the encoder and decoder and use random search for the sizes of the hidden layers. All activations are ReLU, except for the nodes in the code layer and reconstruction layer which have linear activations. We make \edit1{VAEs with two, four, six, and ten latent parameters, stopping at ten} as this is the number of PCA components required to reconstruct non-quasar galaxies in \cite{yip_distributions_2004}. \edit1{The architectures and values for $\lambda$ found by random search are listed in \autoref{tbl:hyperparameters}.}

\begin{table}
\begin{center}
\begin{tabular}{ |c|c|c| }
 \hline
 latent parameters & architecture & $\lambda$ \\
 \hline
 2 & 1000-1663-42-2-42-1663-1000 & 11.2 \\
 \hline
 4 & 1000-1134-64-4-64-1134-1000 & 21.2 \\
 \hline
 6 & 1000-703-94-6-94-703-1000 & 3.02 \\
 \hline
 10 & 1000-549-110-10-110-549-1000 & 7.72\\
 \hline
\end{tabular}
\end{center}
\caption{\edit1{Best architectures and MMD coefficients $\lambda$ found by random search for VAEs with two, four, six, and ten latent parameters.}}
\label{tbl:hyperparameters}
\end{table}

We extract the emission-line equivalent widths and \texttt{spectro1d} spectral classifications from the FITS headers. The objects found by \texttt{spectro1d} to be quasars (\texttt{SPEC\_CLN = 3}) are denoted ``broad-line AGN'' (1.4\%); 83.6\% of these broad-line AGN were targeted as quasars. We subdivide the remaining galaxies (\texttt{SPEC\_CLN = 2}) based on their emission lines. The galaxies with Balmer emission strengths less than $3\sigma$ are denoted ``quiescent galaxies'' (60.1\%), while galaxies with strong Balmer emission are denoted ``emission-line galaxies'' (31.9\%) or ``narrow-line AGN'' (4.0\%) using the AstroML implementation of the \cite{kewley_theoretical_2001} N II/H$\alpha$ line-ratio diagnostic. The code we used to download and process the spectra, train the VAEs, and make the plots in this paper is available at \url{https://github.com/stephenportillo/SDSS-VAE}. Training the VAE, including the hyperparameter search, takes about a day on two cores of an AMD EPYC 7401 processor.

\subsection{Reconstruction Accuracy}

\label{sec:recon}
\begin{figure*}
    \plotone{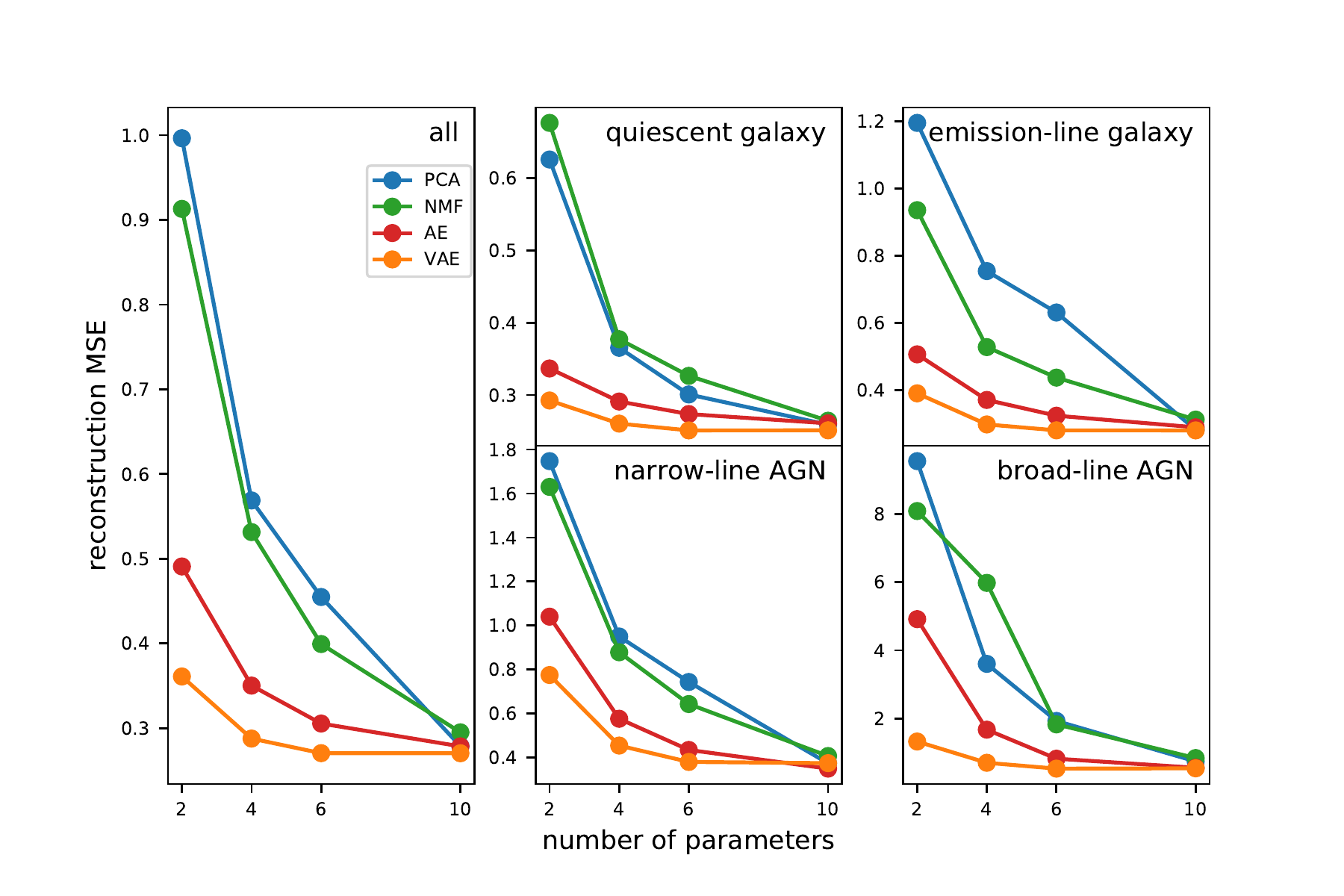}
    \caption{Mean squared reconstruction error for Principal Component Analysis (PCA), non-negative  matrix  factorization (NMF), Non-variational Autoencoder (AE), and VAE. VAE outperforms the other methods, with the greatest advantage being at small numbers of parameters. VAE has the greatest relative advantage for broad-line active galaxies, which also have the highest absolute reconstruction error.}
    \label{fig:reconMSE}
\end{figure*}
To demonstrate VAEs' ability to compress and reconstruct spectra, we train VAEs with two, four, six, and ten latent parameters. We then measure the reconstruction error of these VAEs applied to the validation set of spectra. As a comparison, we create PCA and non-negative matrix factorization (NMF) reconstructions with the same numbers of parameters. Both PCA and NMF reconstruct data as a linear combination of templates. We also train (non-variational) autoencoders (AEs) with the same numbers of latent parameters. \autoref{fig:reconMSE} shows the mean squared reconstruction error of all reconstructions as a function of the number of parameters/components. Generally, at a given number of parameters, the AE and VAE perform better than the two linear methods. NMF gives somewhat better reconstructions than PCA, while the AE gives even better reconstructions than NMF, and the VAE gives the best reconstructions. The performance gaps are most significant when fewer parameters are used. The VAE performs especially well on broad-line AGNs, which have a wide range of emission line widths. The PCA and NMF reconstructions, being linear combinations of templates, capture this variation using templates with deficits at the peaks of spectral lines and excesses in the wings. Therefore, these methods need more components to reconstruct broader spectral lines. By contrast, because the VAE is non-linear, it can control the width of spectral lines with a single parameter. This non-linearity allows a VAE with only two parameters to reconstruct spectra at least as well as a PCA reconstruction with 6 components. The performance gaps between the methods narrow as more parameters are used, but there are still significant differences in the details of the reconstructions.

Increasing the number of VAE parameters from six to ten does not improve the VAE reconstructions much, suggesting that the first six parameters capture more information than the last four. In the ten-parameter VAE, we also find that the latent variances of four of the parameters are typically wider than the prior, meaning that the VAE is not constraining these parameters for each spectrum. This finding indicates that specifying these four parameters is not important for reconstructing spectra, again suggesting that the other six parameters are the most important. We find that the ten parameter VAE sometimes gives slightly worse reconstructions than the six parameter VAE: the four unconstrained parameters seem to be adding noise in these cases.

\begin{figure}
    \plottwo{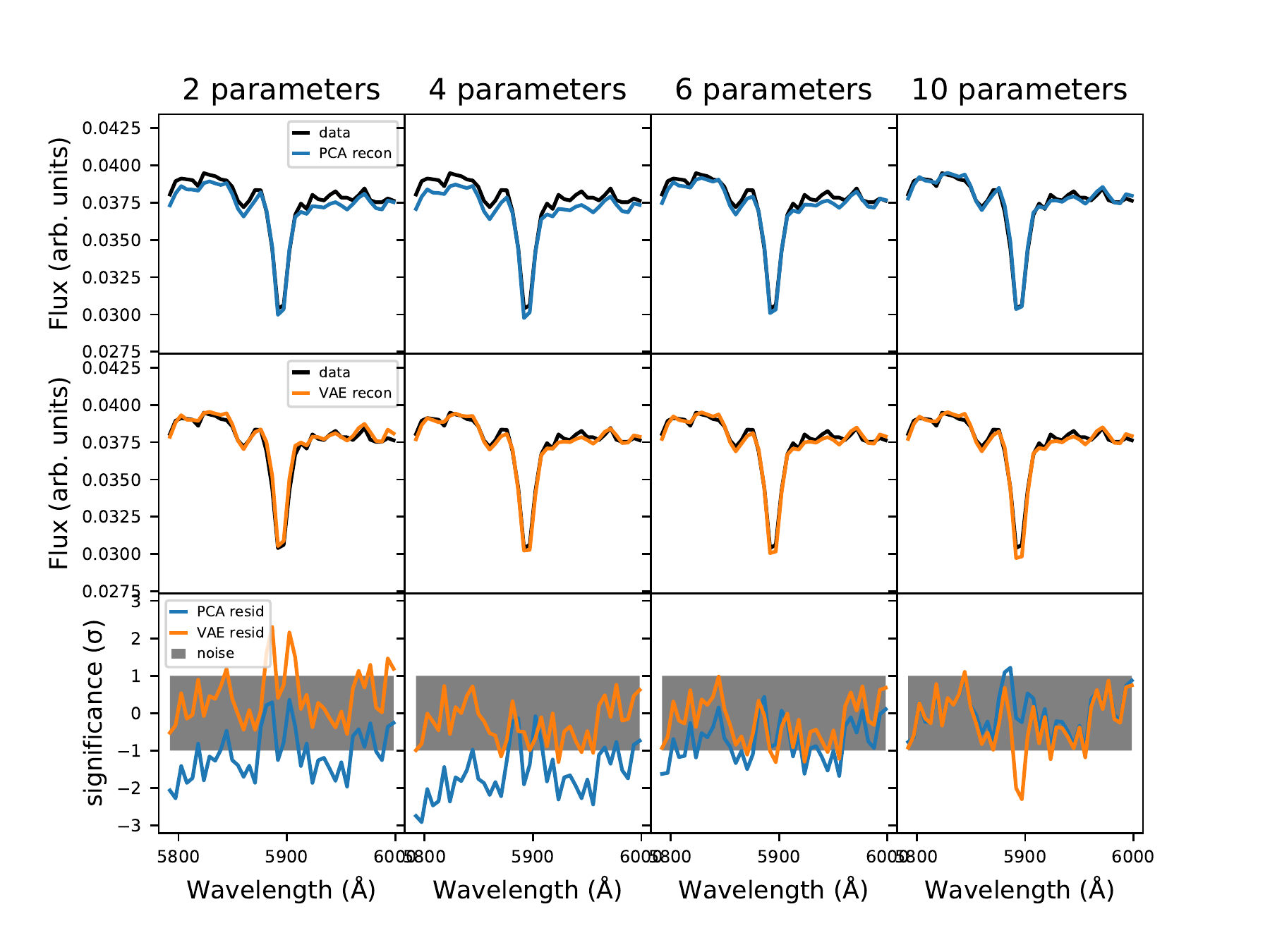}{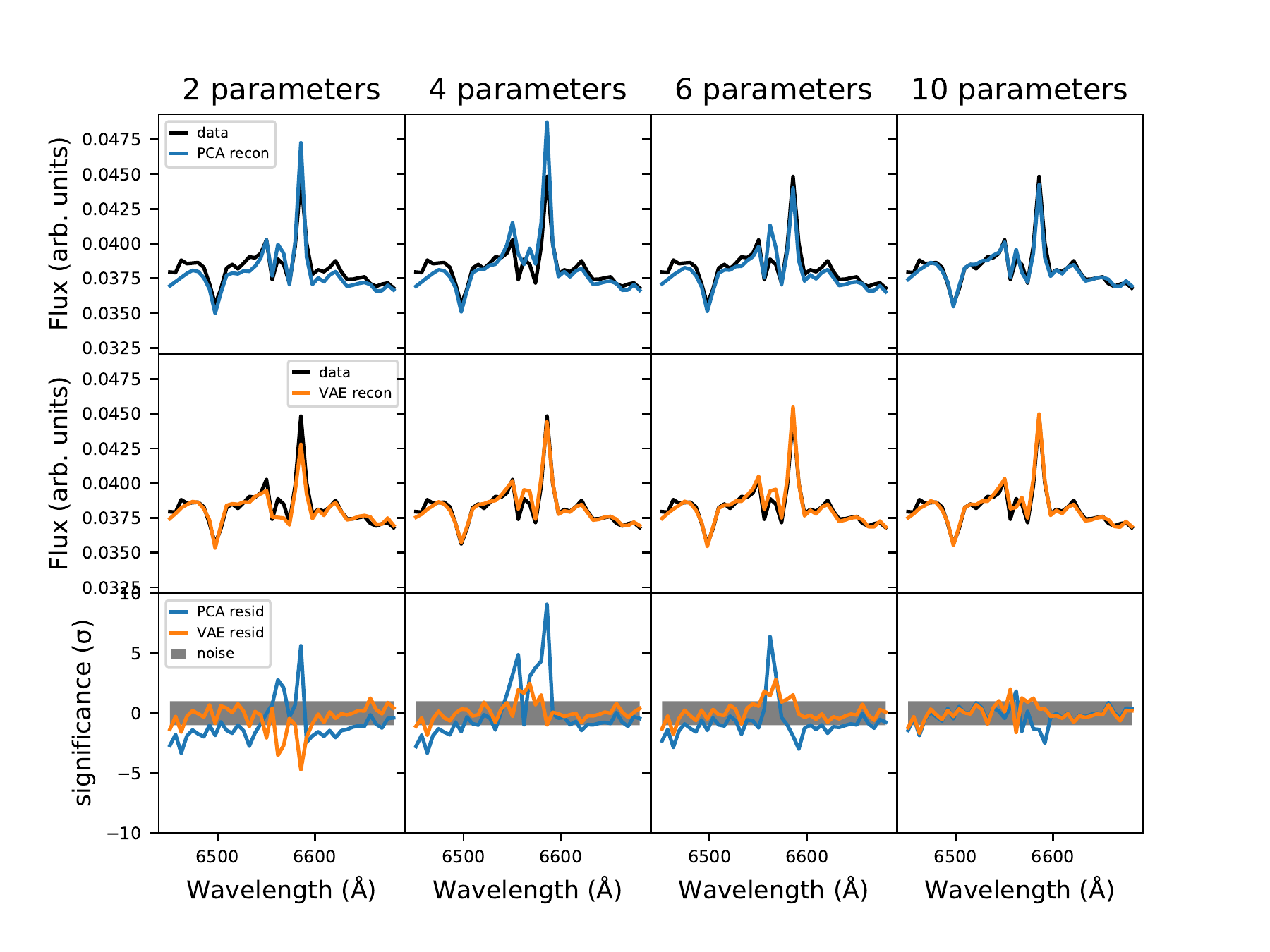}
    \caption{High SNR quiescent galaxy (Plate-MJD-Fiber 394-51913-561): PCA and VAE reconstructions of the Na 5896 absorption line (left) and N II 6550/6585 and H$\alpha$ emission lines (right). The noise band represents the SDSS stated pixel uncertainties. The median SNR of the spectrum is 78 in the left panel and 102 in the right panel. Both reconstructions find the significant Na absorption and N II emission with only two components, but the VAE provides a better reconstruction overall than PCA with the same number of components.}
    \label{fig:recon_hiSNR}
\end{figure}

We now compare the VAE and PCA reconstructions more qualitatively by zooming in on interesting regions of selected spectra in the validation set. We show the reconstruction that corresponds to the latent mean for each spectrum; that is, we do not sample from the latent distribution defined by the latent mean and latent variance, as is done by the VAE during training. We find that variance in the reconstruction that arises from sampling from the latent distribution is much smaller than the spectrum measurement uncertainties. In \autoref{fig:recon_hiSNR}, we show the Na 5896, N II 6550/6585, and H$\alpha$ lines of a high signal to noise quiescent galaxy (median pixel SNR 82). With two parameters, both reconstructions show the Na absorption and N II emission, but the VAE better reproduces the continuum. The PCA reconstruction at 2 parameters shows hints of $H\alpha$ emission that is absent from the VAE reconstruction. Both reconstructions better approximate the amplitude of the lines as more parameters are added, except the VAE reconstruction of the Na absorption lines appears to worsen between six and ten parameters. Still, with four or six parameters, the VAE reconstruction better fits the continuum and the lines.

\begin{figure}
    \plottwo{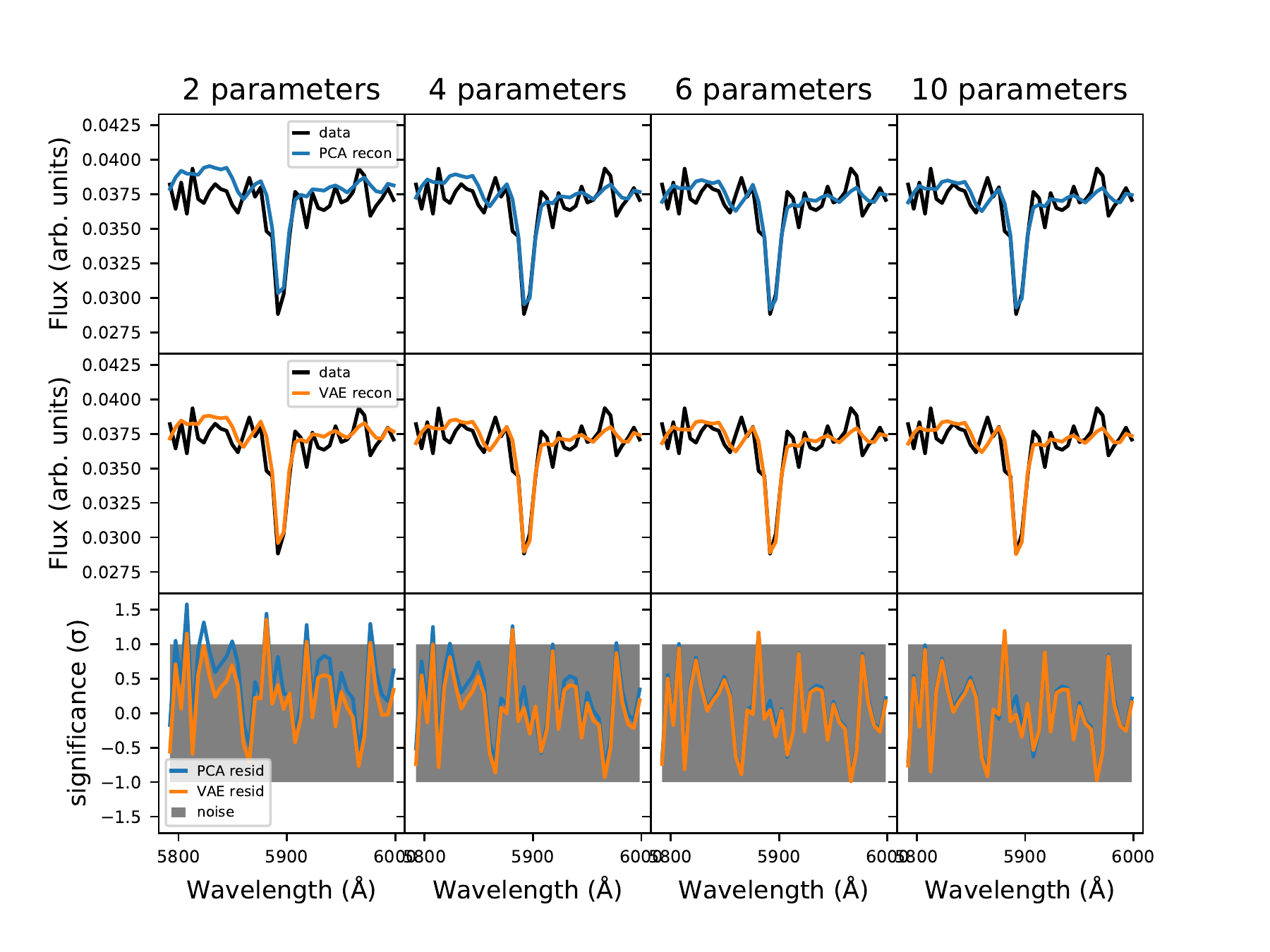}{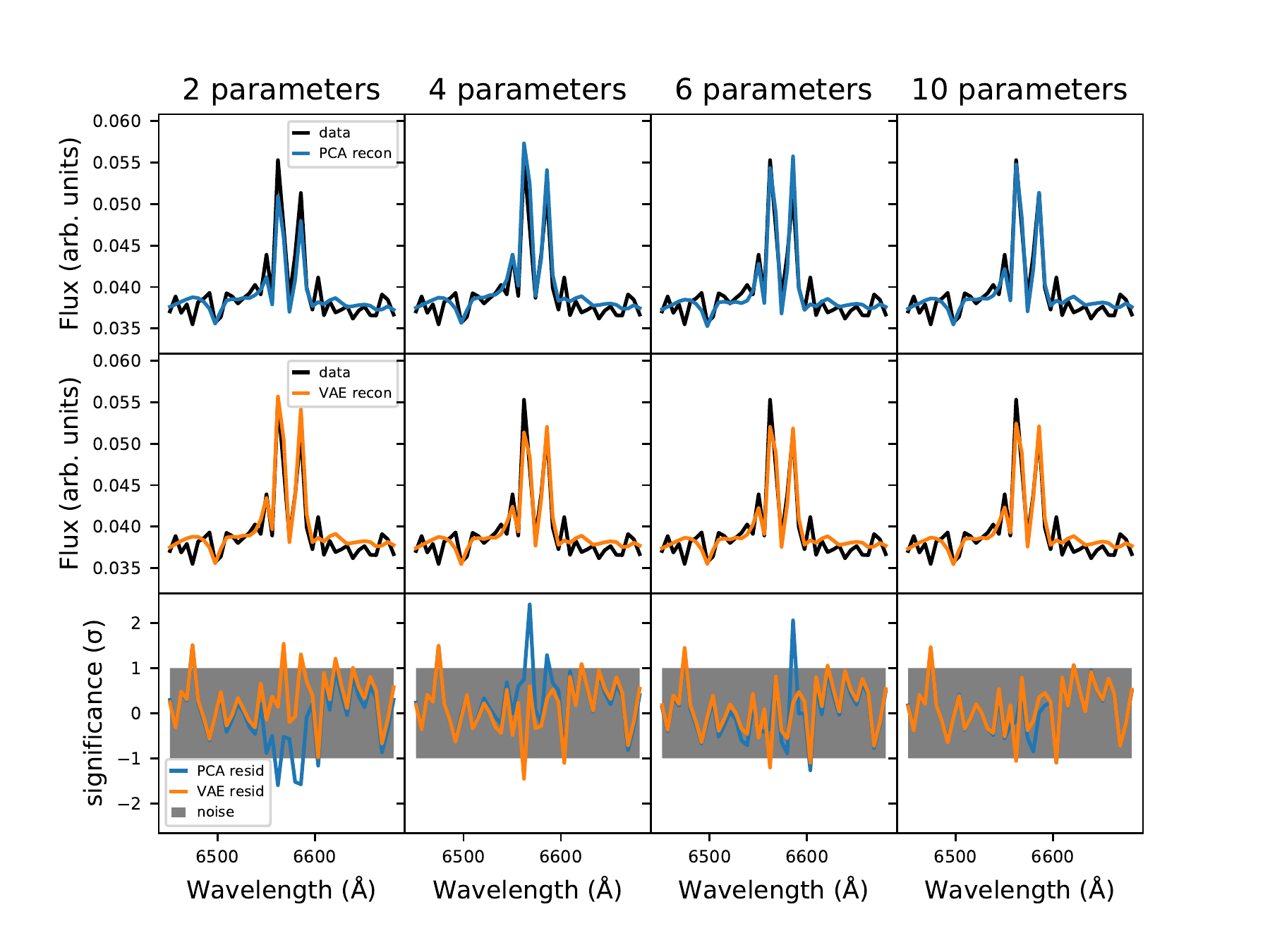}
    \caption{Low SNR quiescent galaxy (Plate-MJD-Fiber 497-51989-573): PCA and VAE reconstructions of the Na 5896 absorption line (left) and N II 6550/6585 and H$\alpha$ emission lines (right). The median SNR of the spectrum is 12 in the left panel and 18 in the right panel. Even at low SNR, both reconstructions find the Na absorption, H$\alpha$, and N II emission with only two components. The VAE still provides a better reconstruction compared to PCA with the same number of parameters, although the differences are less significant at low SNR.}
    \label{fig:recon_loSNR}
\end{figure}

In \autoref{fig:recon_loSNR}, we show the Na 5896, N II 6550/6585, and H$\alpha$ lines of a low signal to noise quiescent galaxy (median pixel SNR 11). Even with the lower SNR, both reconstructions capture the Na absorption, N II emission, and $H\alpha$ emission with only two components. The differences between the two reconstructions are less pronounced at this lower SNR. At four and six components, the PCA reconstruction overestimates the flux in the H$\alpha$ line and N II 6565 line, respectively, but these residuals do not exceed 2$\sigma$.

\begin{figure}
    \plottwo{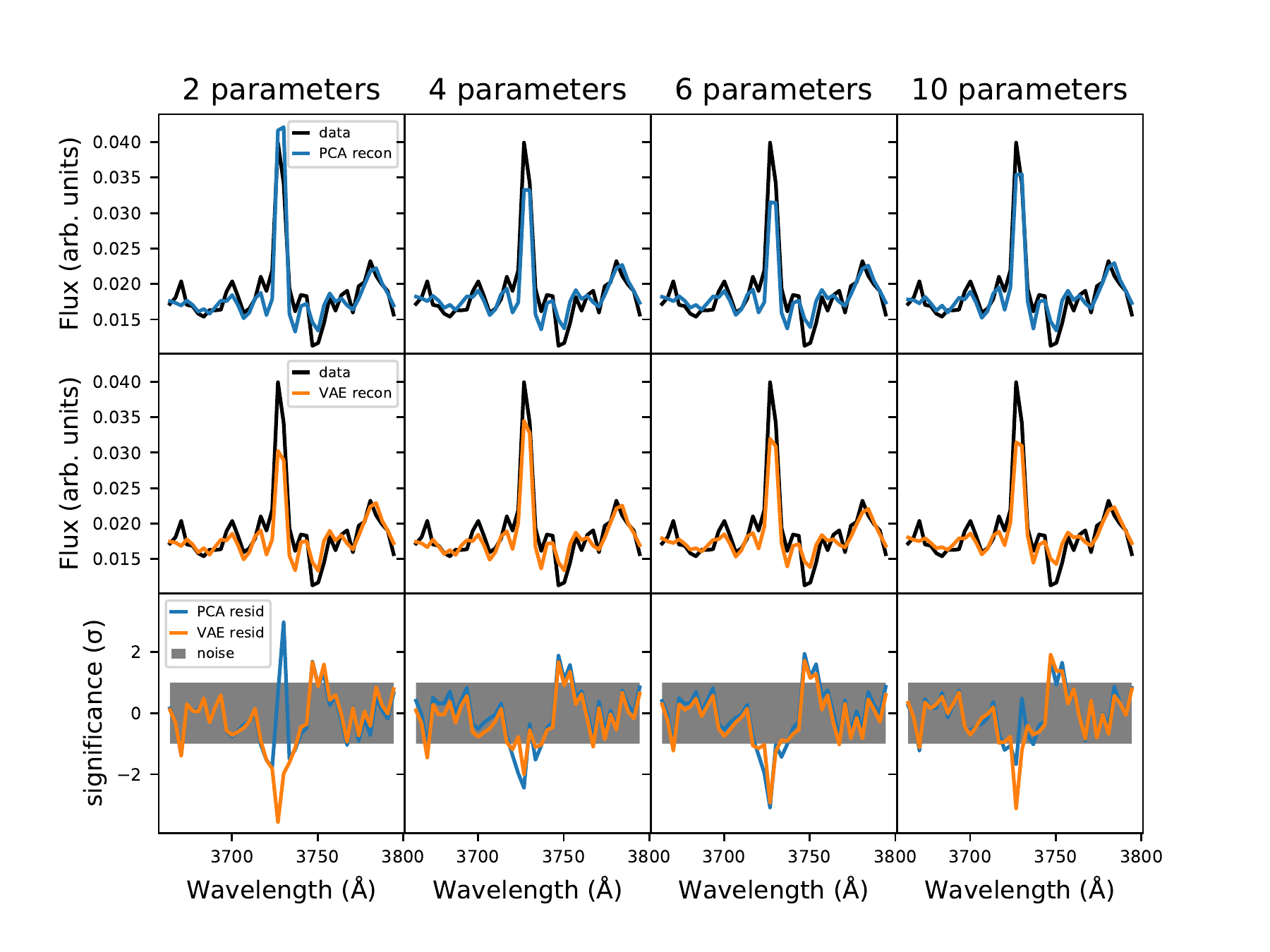}{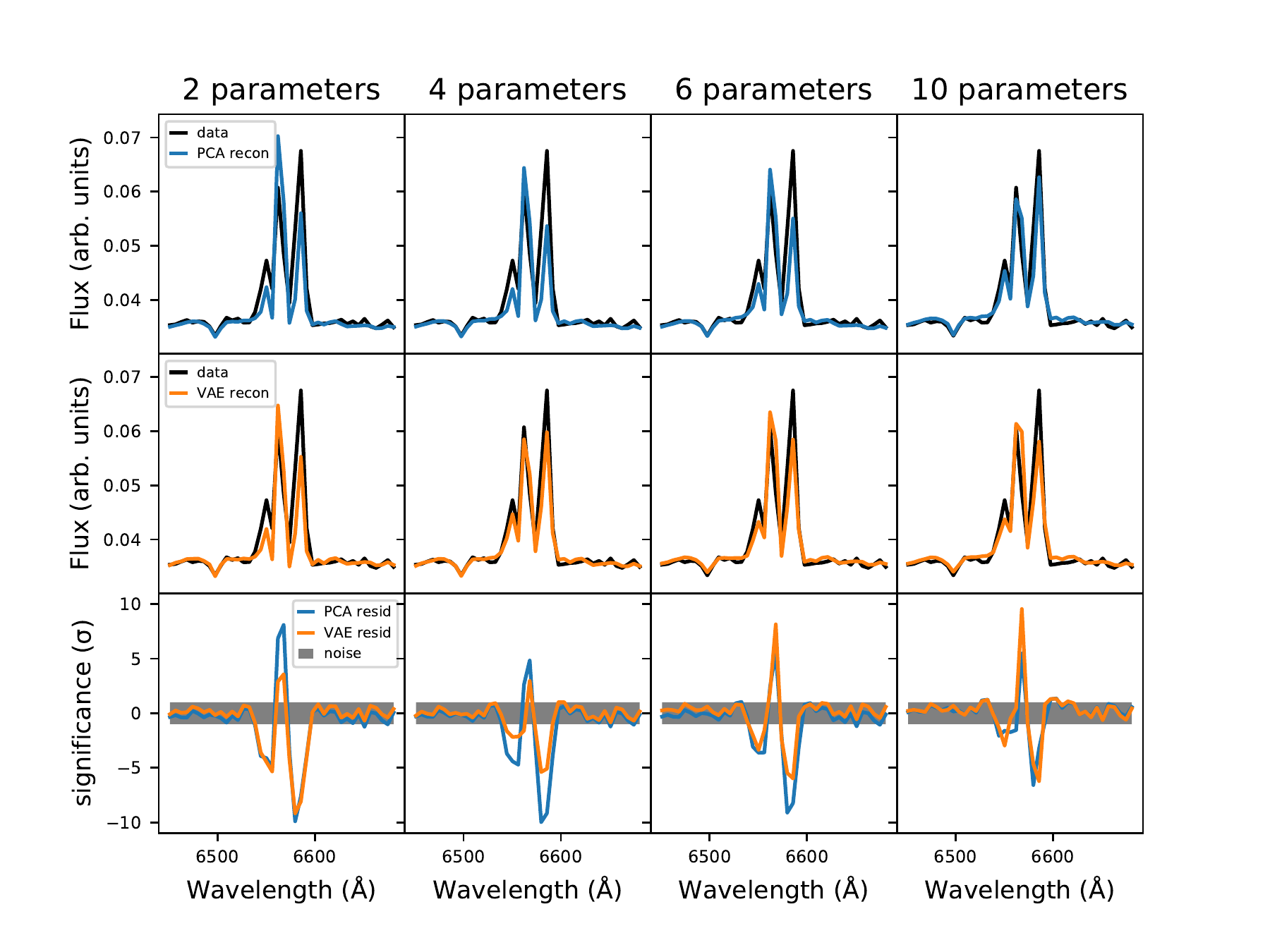}
    \caption{Extreme emission-line galaxy (Plate-MJD-Fiber 391-51782-353): PCA and VAE reconstructions of the O II 3726/3729 (left) and H$\alpha$ and S II emission lines (right). The median SNR of the spectrum is 9 in the left panel and 13 in the right panel.  While the PCA reconstruction of the strong O II 3727 line gets better with more components, the VAE reconstruction does not improve after four parameters.}
    \label{fig:recon_EL}
\end{figure}

In \autoref{fig:recon_EL}, we show the O II 3727/3730, H$\alpha$, and N II 6550/6585 lines of an emission-line galaxy. The lines have equivalent widths of 22, 25, 62, 2, and 7 {\AA}, respectively. While all the lines are present in both reconstructions with only two parameters, both reconstructions struggle to match the amplitudes of the lines. With four parameters or more, both reconstructions underestimate the O II and N II emission while overestimating the H$\alpha$ emission. The VAE reconstruction does not improve after four parameters, while the PCA reconstruction does improve with ten parameters. \edit1{This mismatch may be due to our choice to cap the pixel SNR to 50 (see the beginning of \autoref{sec:application}), which puts less emphasis on reconstructing high SNR pixels like extreme emission lines.}

\begin{figure}
    \plottwo{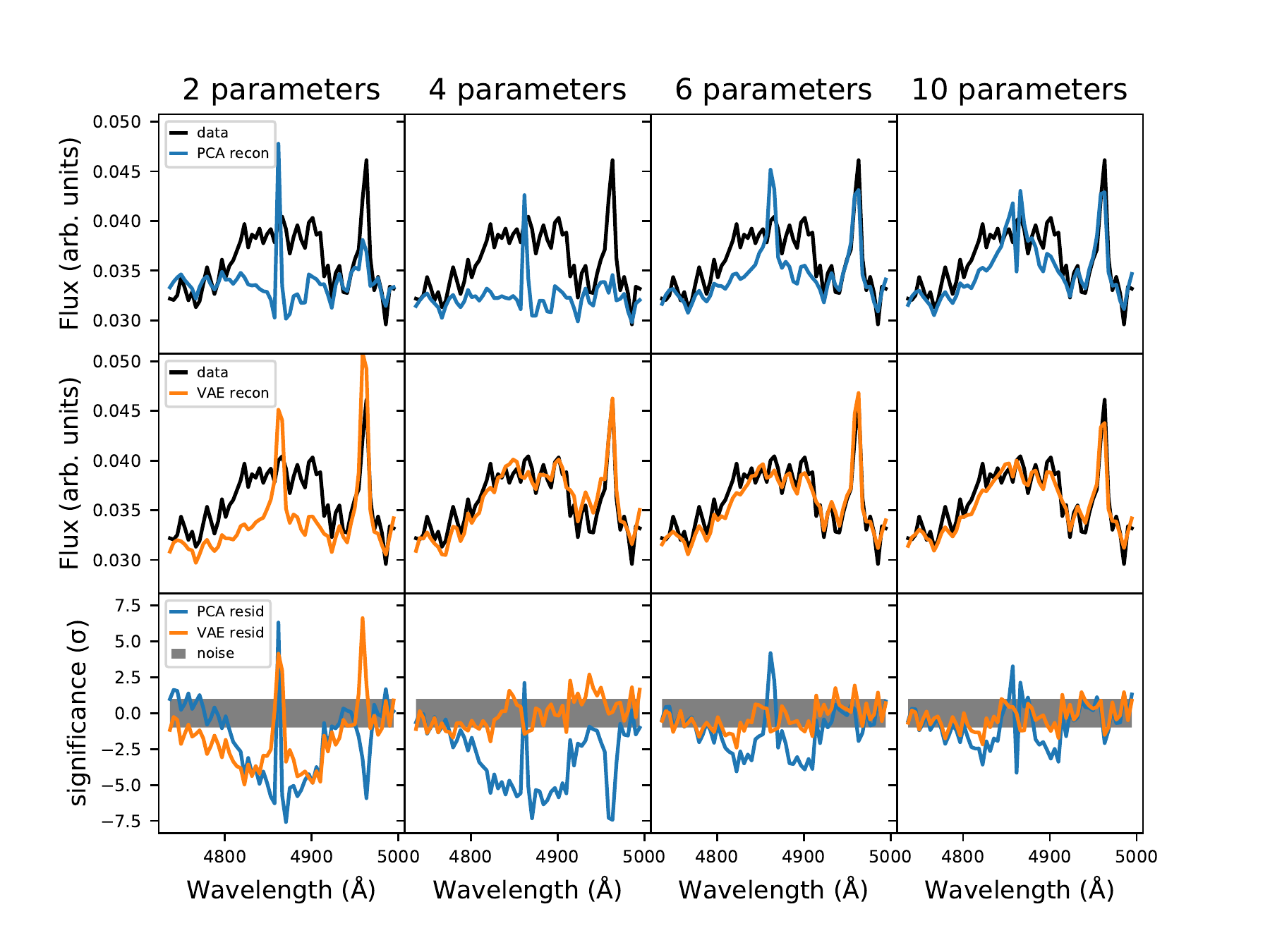}{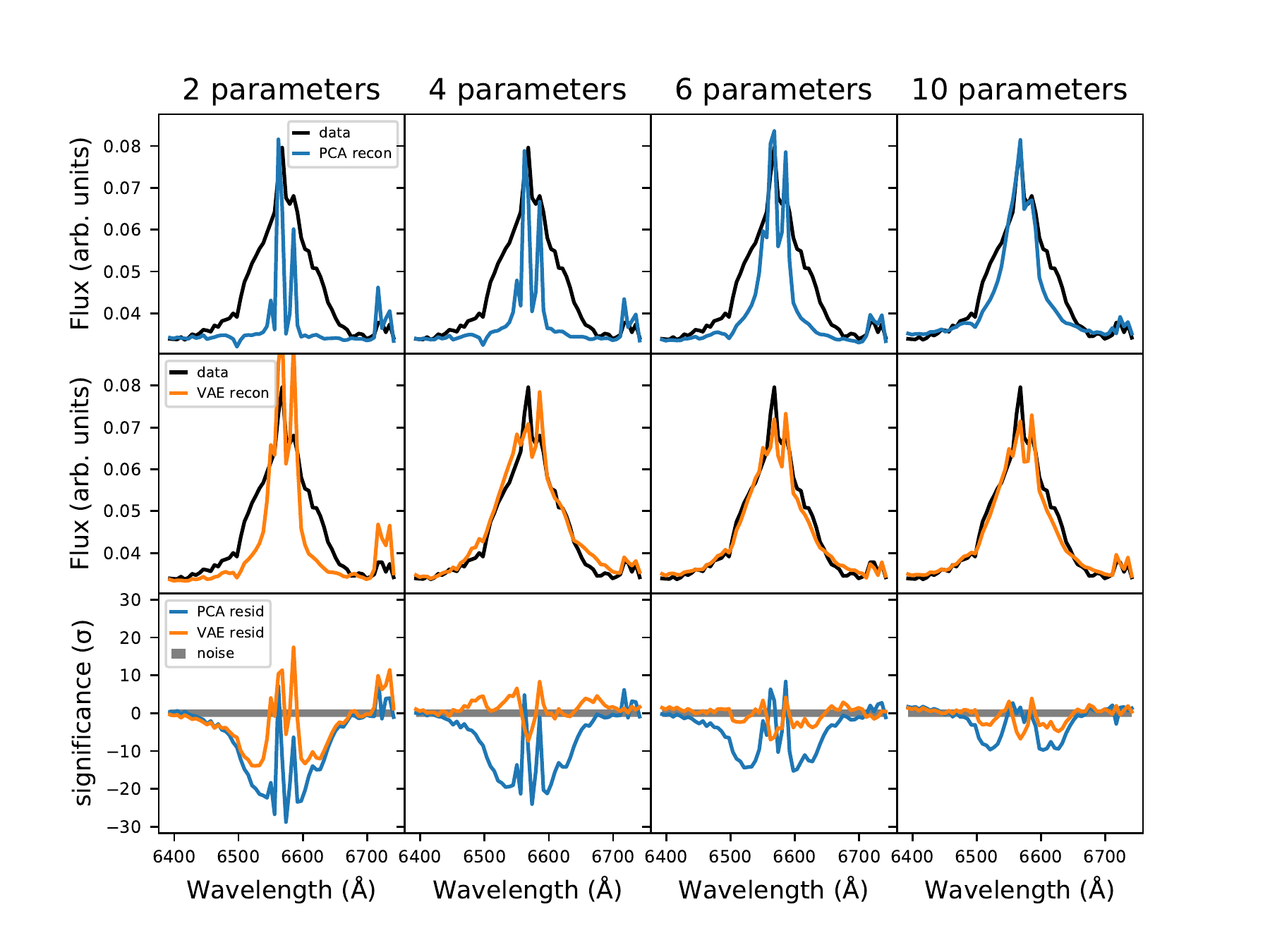}
    \caption{Broad-line active galaxy (Plate-MJD-Fiber 541-51959-519): PCA and VAE reconstructions of the H$\beta$ and O III 4959 emission lines (left) and H$\alpha$, N II 6550/6585, and S II 6718/6733 emission lines (right). The median SNR of the spectrum is 30 in the left panel and 47 in the right panel. The VAE reconstructs the amplitude and width of the lines with four parameters, while the lines in the PCA reconstruction are too narrow, even with ten parameters.}
    \label{fig:recon_BL}
\end{figure}

In \autoref{fig:recon_BL}, we show the H$\beta$, O III 4959, H$\alpha$, N II 6550/6585, and S II 6718/6733 lines of a broad-line active galaxy. The SDSS pipeline reports that the Balmer lines have a width of $\approx 3000$ km/s while the other lines have a width of $\approx 1800$ km/s. The four-parameter VAE captures the width of all of these spectral lines well. The four-component PCA reconstruction has these emission lines, but they are far too narrow. As more components are added, the PCA reconstruction improves, but even with ten components, the emission lines are not wide enough.

\subsection{Interpretation of VAE Tracks}
\label{sec:VAEtracks}

\begin{figure*}
    \plotone{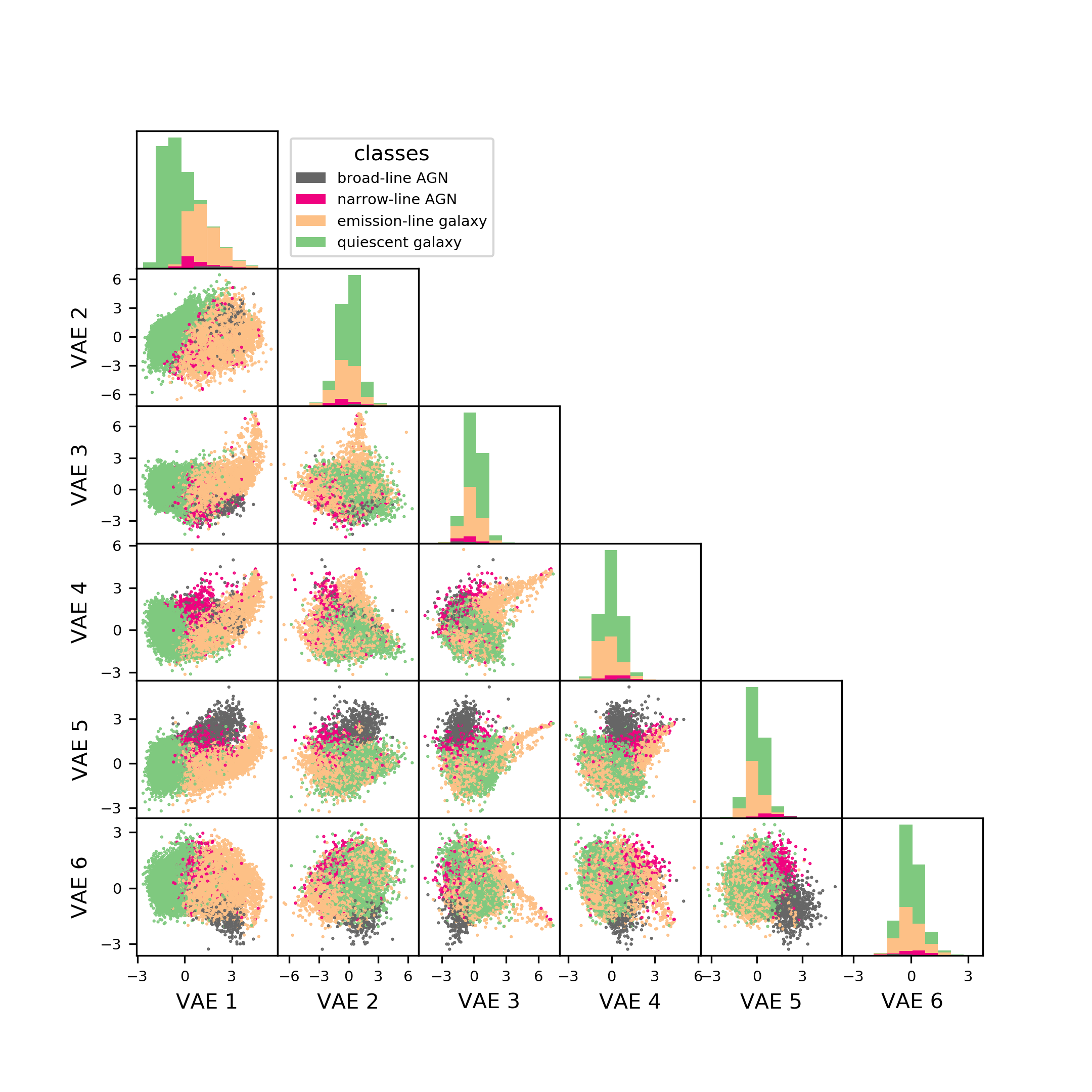}
    \caption{Corner plot of the first six VAE components of all training and validation spectra. The spectra are color-coded based on the classification outlined in \autoref{sec:application}.}
    \label{fig:cornerplot}
\end{figure*}

In \autoref{fig:cornerplot}, we show a corner plot of the latent mean of each galaxy in VAE latent space. The axes of the VAE latent space are arbitrary and may not be the most human-interpretable directions in latent space. We construct a basis in latent space by using PCA on the latent means to find vectors that describe the variance of the spectra \textit{in latent space}. This procedure differs from using PCA directly on the spectra, which finds eigenspectra that describe the variance of the spectra \textit{in spectrum space}. The last four components explain very little of the variance in latent space, which is expected since the six-dimensional VAE can reconstruct spectra nearly as well as the ten-dimensional VAE, as discussed in \autoref{sec:recon}. Thus, we drop the last four components and only plot projections onto the first six components. For brevity, we will refer to these PCA components of the VAE latent space simply as VAE 1 through VAE 6, in descending order of explained variance.

The VAE latent space separates quiescent galaxies, emission-line galaxies, narrow-line AGN, and broad-line AGN. This separation can most clearly be seen in the scatter plot of VAE 1 and VAE 5 in \autoref{fig:cornerplot}. This projection shows two distinct arms: one that goes from quiescent galaxies to emission-line galaxies, and another that splits off for the narrow-line and broad-line AGN. The hook at the end of the emission-line track corresponds to the extended linear feature that can be seen in many of the other projections.

\begin{figure*}
    \plotone{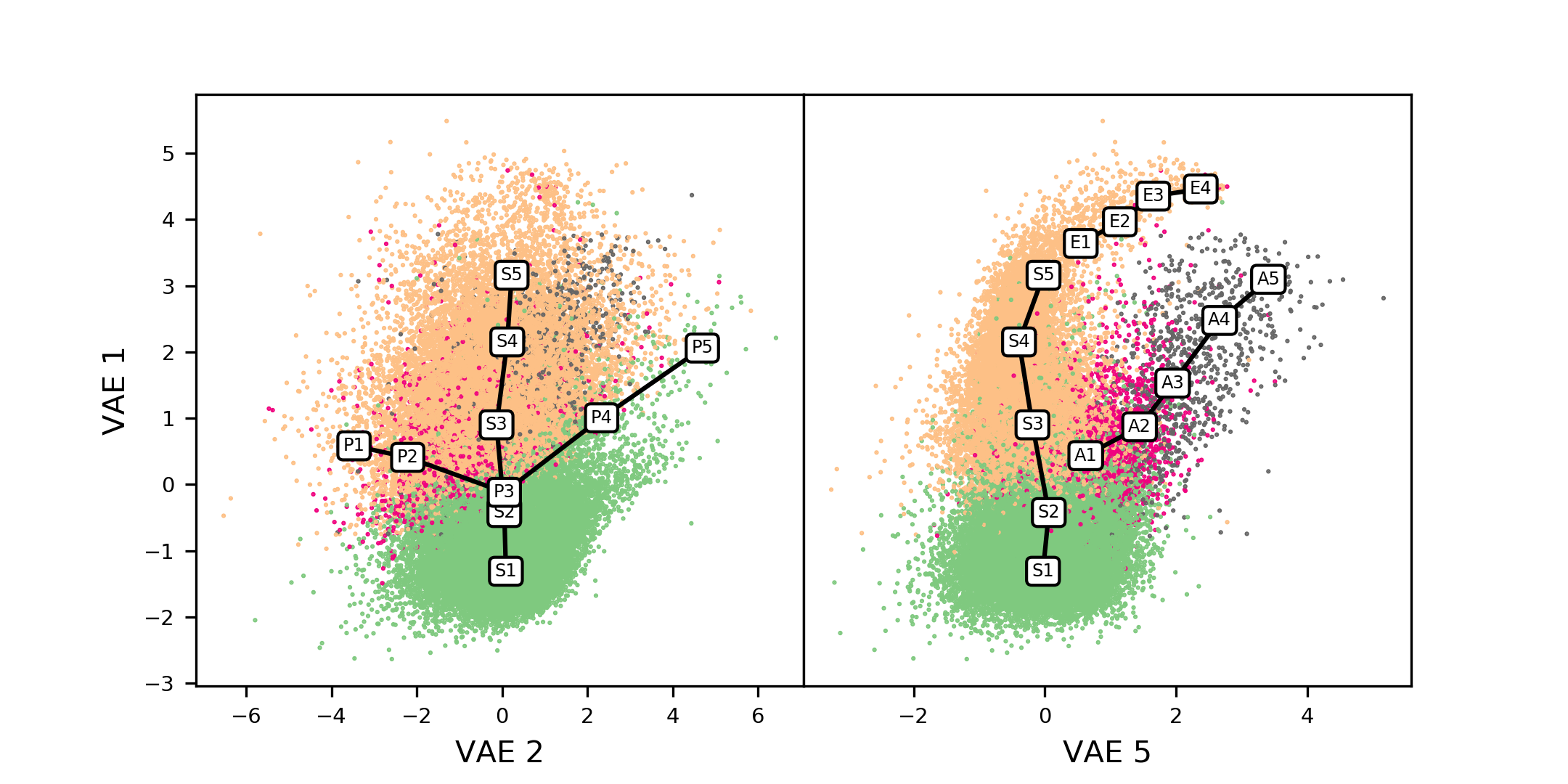}
    \caption{Scatter plot of the first, second, and fifth VAE components, with the four tracks discussed in \autoref{sec:VAEtracks} overlaid. \edit1{The tracks are the (S) star formation track, the (E) extreme line emitters, the (P) post-starburst track, and the (A) active galaxy track. }}
    \label{fig:ztracks}
\end{figure*}

To interpret the VAE latent space, we create tracks that follow the distribution of galaxies and then use the VAE to generate synthetic spectra along these tracks. To create a track, we first choose some spectral classes to follow and a VAE component (or linear combination of components) to use to bin the spectra. In each bin, we find the centroid of all included spectra that fall into this bin and then make this centroid a point in the track. For example, to create the star formation track, we first select the quiescent and emission-line spectra and then bin these spectra by their VAE 1 values. Then for each bin in VAE 1, we find the centroid in latent space and add that point to the track.  We show these tracks in two projections (VAE 1 vs. VAE 2 and VAE 1 vs VAE 5) in \autoref{fig:ztracks}. These sequences of spectra show how galaxy spectra change as we move in the VAE latent space.

\subsubsection{Star Formation Track}
\begin{figure}
    \plotone{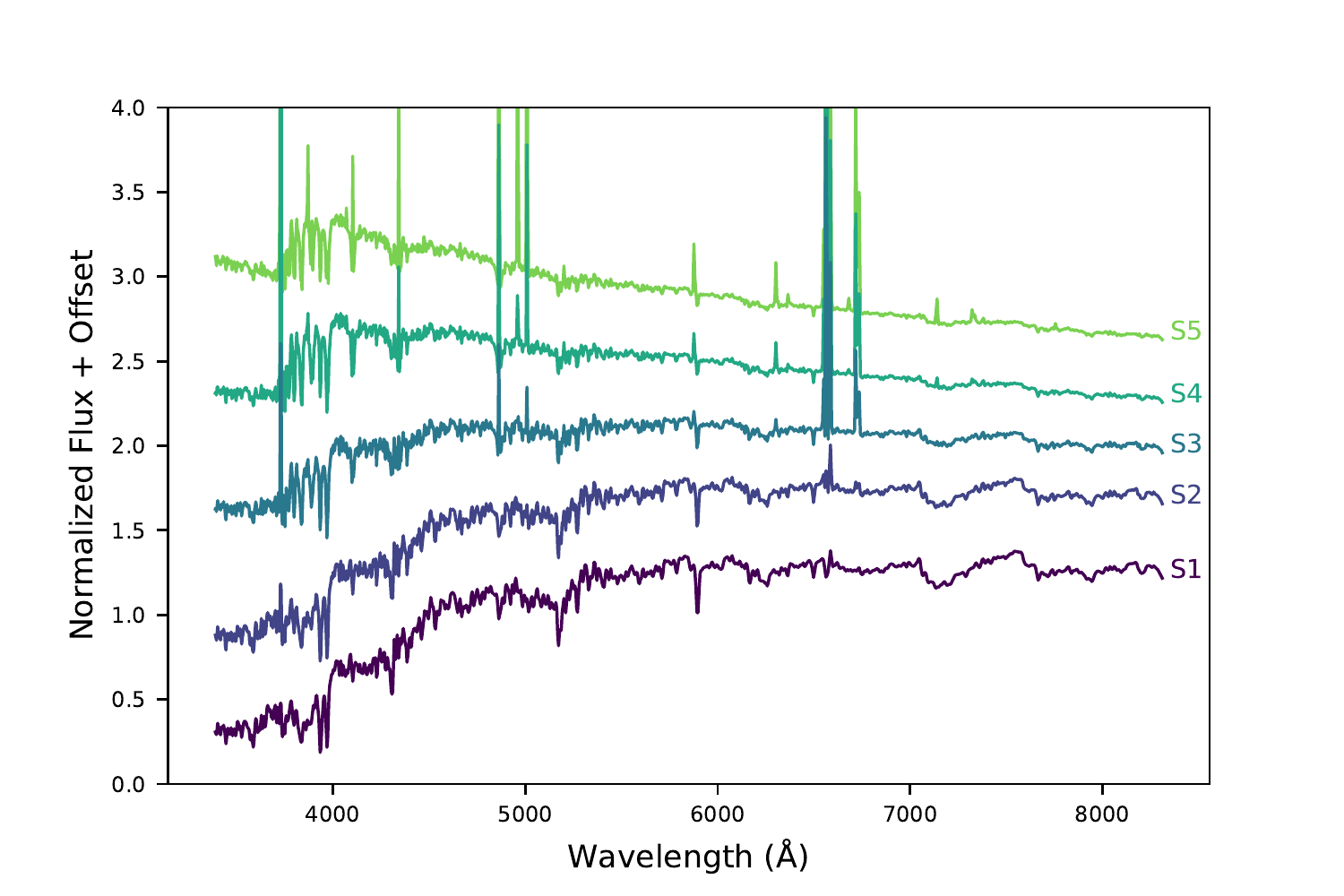}
    \caption{Synthetic spectra along the star formation track in \autoref{fig:cornerplot}, plotted with offsets added to separate the spectra. Proceeding through the sequence, the continuum changes from that of an old stellar population to a young stellar population, and nebular emission lines appear and strengthen.}
    \label{fig:SFtrack}
\end{figure}

The first track we create goes from quiescent galaxies to emission-line galaxies (labeled S1-5 in \autoref{fig:ztracks}), and its spectra are plotted in \autoref{fig:SFtrack}. This track mostly follows VAE 1, and we stop it before it goes into the emission-line galaxy hook that is seen in the projection of the VAE 1 and VAE 5. This track mostly changes in VAE 1 without moving in the other components. At S1, the spectrum is of a quiescent galaxy: the continuum is red, indicating an old stellar population, with absorption lines like Ca H, Ca K, G band,  Mg, and Na. As we proceed towards S5, the continuum gets bluer, indicating younger stellar populations, and nebular emission lines start to appear. First at S2, hints of OII 3726/3729 and NII 6583 appear as VAE 1 increases. Then these emission lines strengthen and are joined by lines like S II, OIII 5007, and the Balmer series. This track acts similarly to the second PCA component in \cite{yip_distributions_2004}: red galaxies have large, positive second PCA coefficients while blue galaxies have small, or even negative, second PCA coefficients.

\subsubsection{Extreme Line Emitters}
\begin{figure}
    \plotone{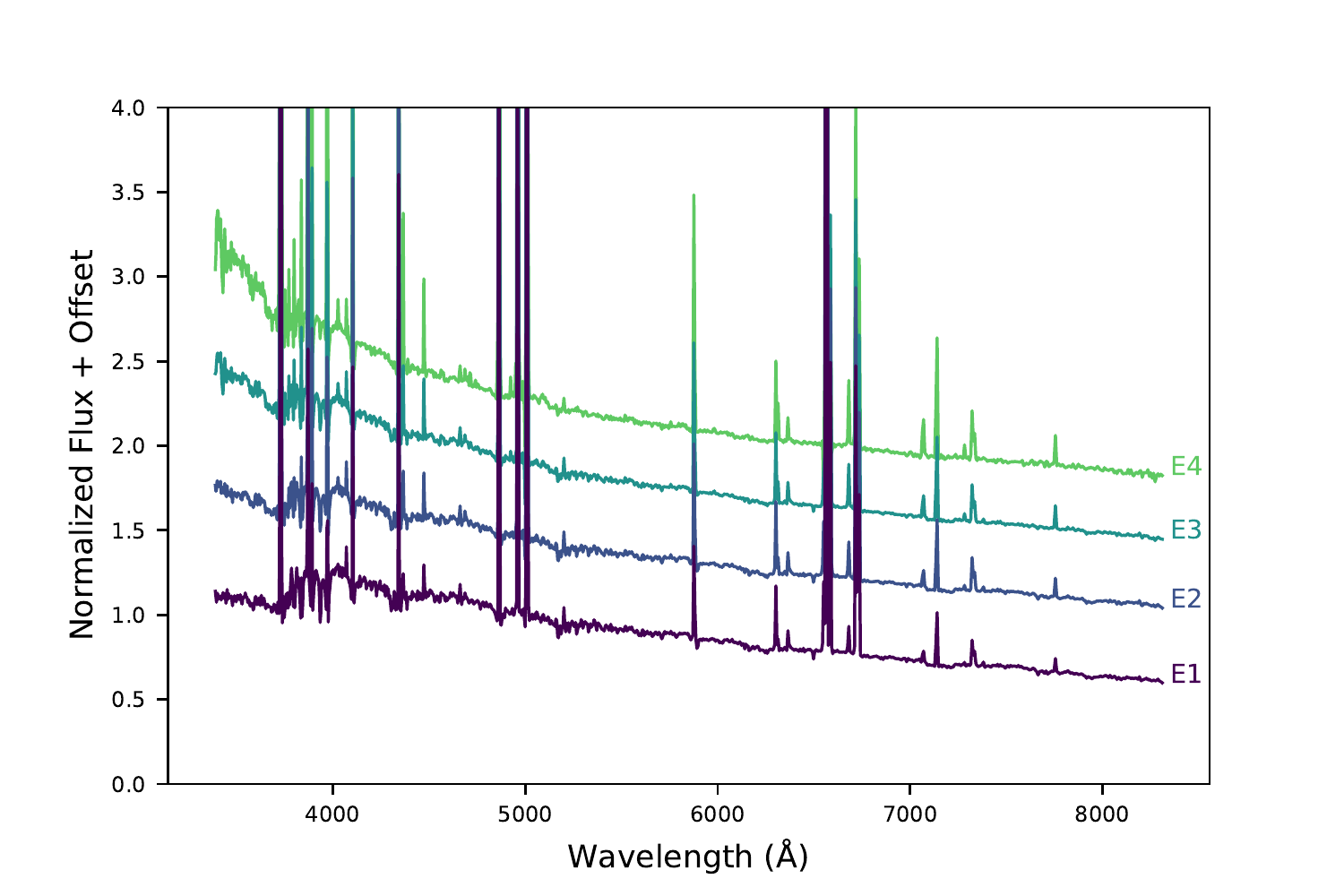}
    \caption{Synthetic spectra along the starburst sequence in \autoref{fig:cornerplot}, plotted with offsets added to separate the spectra. Proceeding through the sequence, the nebular emission lines strengthen and the blue end of the continuum rises.}
    \label{fig:ELtrack}
\end{figure}

Our second track (labeled E1-4 in \autoref{fig:ztracks}) picks up at the end of the previous track and continues into the emission-line galaxy hook seen in VAE 1 and VAE 5. This track's spectra are plotted in \autoref{fig:ELtrack}. These galaxies all have extremely strong line emission, and as the track continues up the hook, the emission lines get stronger and the continuum blueward of 4000 {\AA} increases. From E1 to E4, log(O III/H$\beta$) stays almost constant at $\approx 0.3$, while log(S II/H$\alpha$) decreases from -0.5 to -0.7 and log(N II/H$\alpha$) decreases from -0.7 to -1.6. \cite{yip_distributions_2004} find a similar track in their parameters $(\phi_{\rm KL}, \theta_{\rm KL})$, which are transformations of their first three PCA coefficients. They find a track in this space that proceeds from red to blue to extreme emission-line galaxies, showing a similar progression as our S and E tracks put together.

\subsubsection{Post-starburst Track}
\begin{figure}
    \plotone{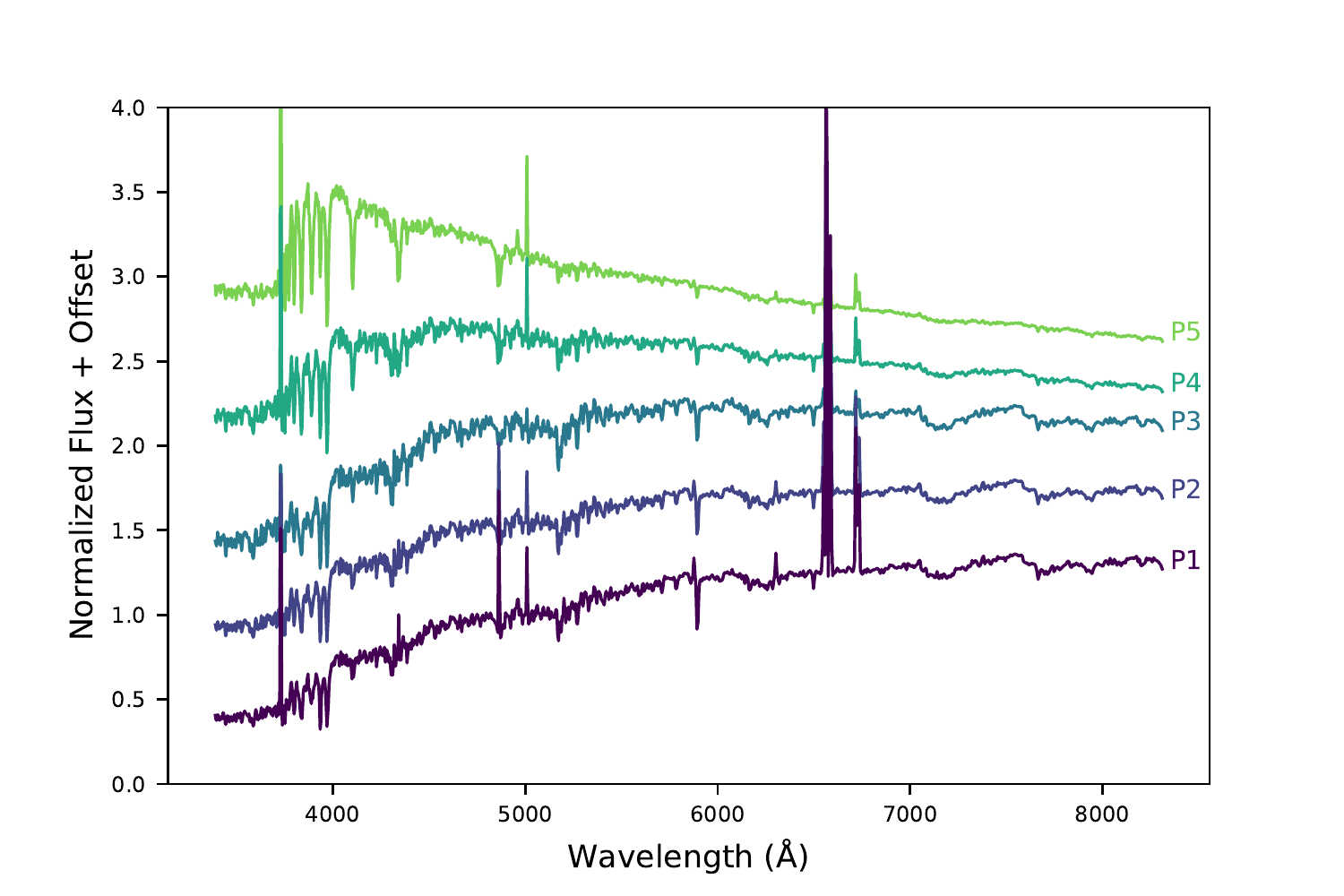}
    \caption{Synthetic spectra along the post-starburst sequence in \autoref{fig:cornerplot}, plotted with offsets added to separate the spectra. The sequence starts with an old stellar population with nebular lines and ends with post-starburst spectrum with a young stellar population but weak nebular emission lines.}
    \label{fig:SBtrack}
\end{figure}
The third track we create follows the quiescent galaxies as VAE 2 changes (labeled P1-5 in \autoref{fig:ztracks}), with the spectra plotted in \autoref{fig:SBtrack}. The middle of this track (P3) intersects the star formation track near S2. At P1, the spectrum has a red continuum, indicating older stellar populations, but also has nebular emission lines. Proceeding along the track, the continuum gets bluer and the nebular emission lines weaken. P4 and P5 are post-starburst spectra, with a blue continuum from young stars and weak nebular emission lines. Similarly, \cite{yip_distributions_2004} find that galaxies with negative second and third PCA coefficients are post-starburst galaxies.

\subsubsection{Active Galaxy Track}
\begin{figure}
    \plotone{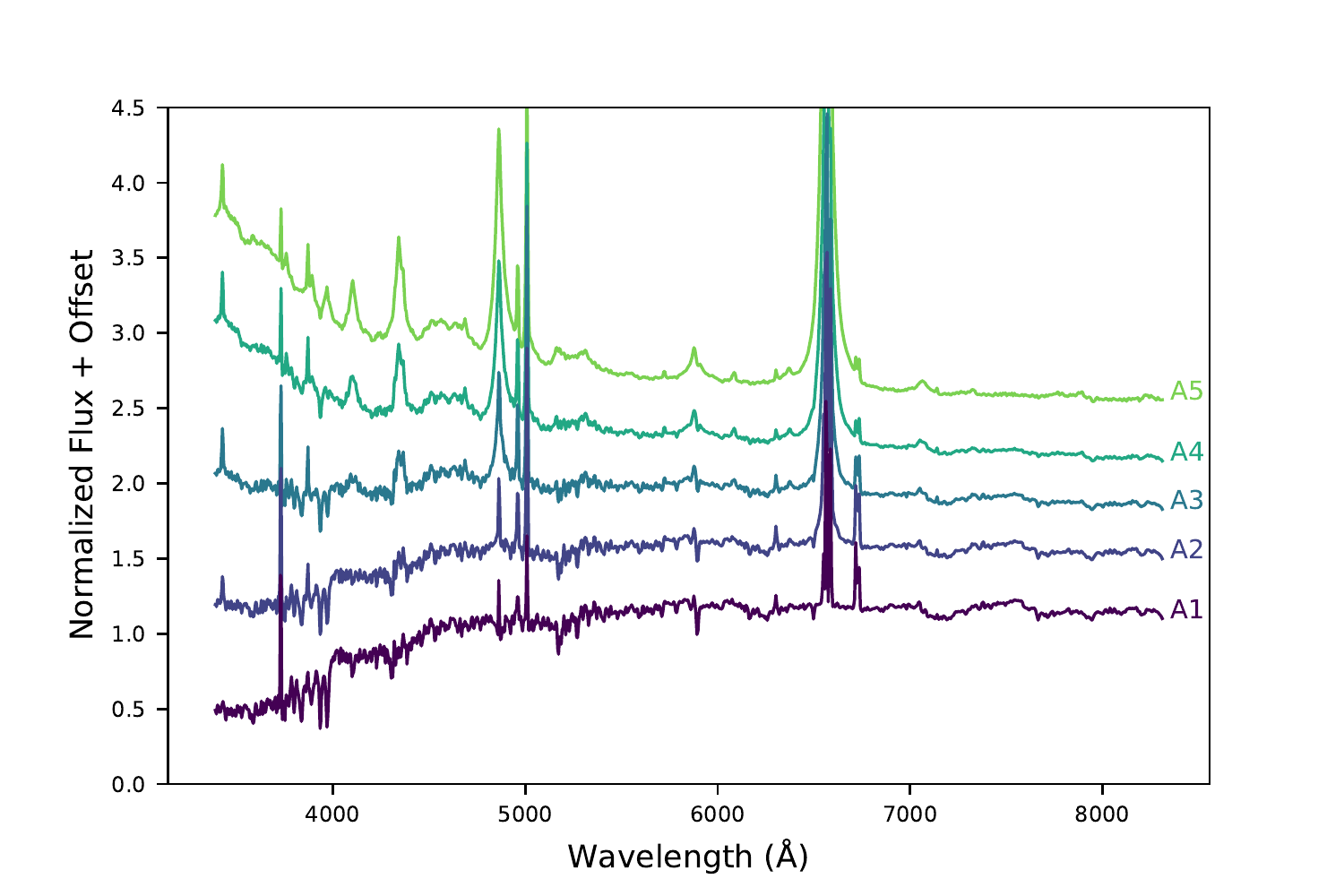}
    \caption{Synthetic spectra along the active galaxy sequence in \autoref{fig:cornerplot}, plotted with offsets added to separate the spectra. The sequence progresses from a Type 2 Seyfert spectrum with only narrow emission lines to a Type 1 Seyfert spectrum with broad Balmer emission lines.}
    \label{fig:AGNtrack}
\end{figure}

This track (labeled A1-5 in \autoref{fig:ztracks}) starts near S3 on the star formation track and diverges upward in VAE 5 as VAE 1 increases. The spectra along this track are plotted in \autoref{fig:AGNtrack}. At A1, the spectrum has narrow-emission lines, while at A5, the spectrum is a broad-line active galaxy. Proceeding along the track, the nebular emission lines get stronger and the Balmer lines get stronger and broaden. The continuum also gets bluer, especially bluewards of 4000 {\AA}.
\subsubsection{Individual VAE Components}
\label{sec:VAEtrackcomp}

\begin{figure*}
    \plotone{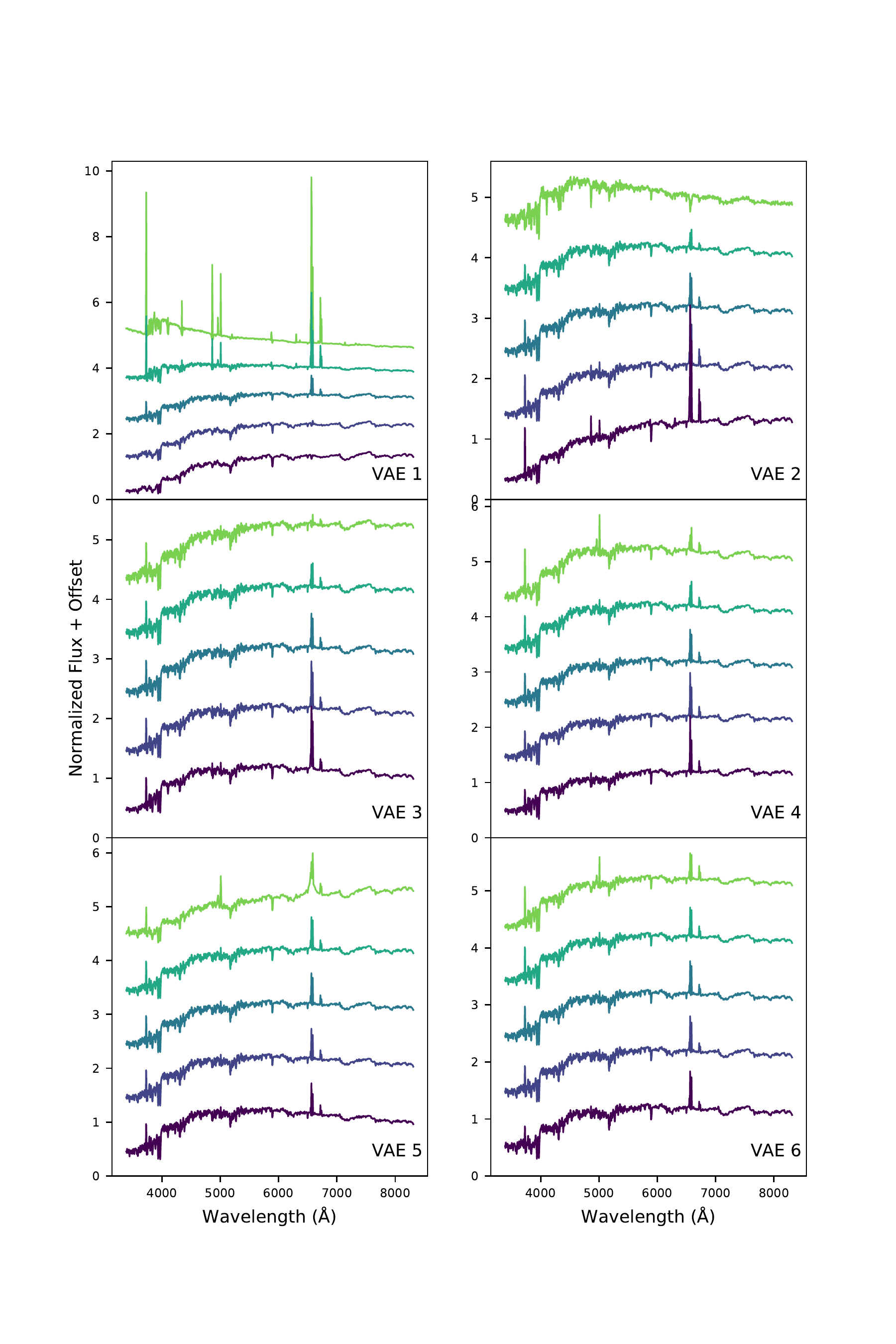}
    \caption{Synthetic spectra along sequences where one VAE parameter is changed at a time, plotted with offsets added to separate the spectra. The middle spectrum in each sequence corresponds to the centroid of the VAE latent means of all spectra. These sequences are discussed in \autoref{sec:VAEtrackcomp}.}
    \label{fig:PCAtrack}
\end{figure*}

We also create tracks that change only one VAE component at a time, with the middle of each track intersecting with the centroid of all galaxies' latent means. The spectra from these tracks are shown in \autoref{fig:PCAtrack}. These spectra show the effect that a VAE component has in isolation, but do not show effects that happen when multiple components are changed from their mean values.

Increasing VAE 1 changes spectra in a similar way as going along the star formation track: strengthening the nebular emission lines and making the continuum bluer. This behavior is expected, as the main component that changes along the star formation track is VAE 1. By contrast, decreasing VAE 2 strengthens the nebular emission lines while making the continuum slightly redder. OII 3727 and OIII are also more prominent as VAE 2 decreases. These two parameters give the VAE the ability to reconstruct different combinations of stellar population age and star formation activity. Increasing VAE 1 corresponds to a younger stellar population and higher star formation activity, while increasing VAE 2 corresponds to a younger stellar population and lower star formation activity. Given this combination of effects, it is not suprising that the post-starburst galaxies (P4 and P5) have high VAE 1 and VAE 2.

Increasing VAE 3 weakens H$\alpha$ and NII while retaining OII 3727. The continuum does not change much, in contrast to changing VAE 1 and VAE 2. Increasing VAE 4 weakens H$\alpha$ and NII while also strengthening OII 3727 and OIII 5007 and making the spectrum slightly bluer. 

Increasing VAE 5 broadens the H$\alpha$ line and also makes the continuum redder. Increasing VAE 6 weakens H$\alpha$ and NII while strengthening OII 3727 and OIII 5007, similarly to VAE 4. Unlike increasing VAE 4, increasing VAE 6 does not make the continuum bluer.

The six VAE components give the VAE the flexibility to fit different combinations of continuum color and emission line strengths. While these sequences for each VAE component can be interpreted, we must emphasize that these sequences were all chosen to intersect at the centroid of latent space. Because the VAE is non-linear, the effect of each component can vary as a function of position within latent space. 

\subsection{Outlier Identification}
\label{sec:outliers}
\begin{figure}
    \plotone{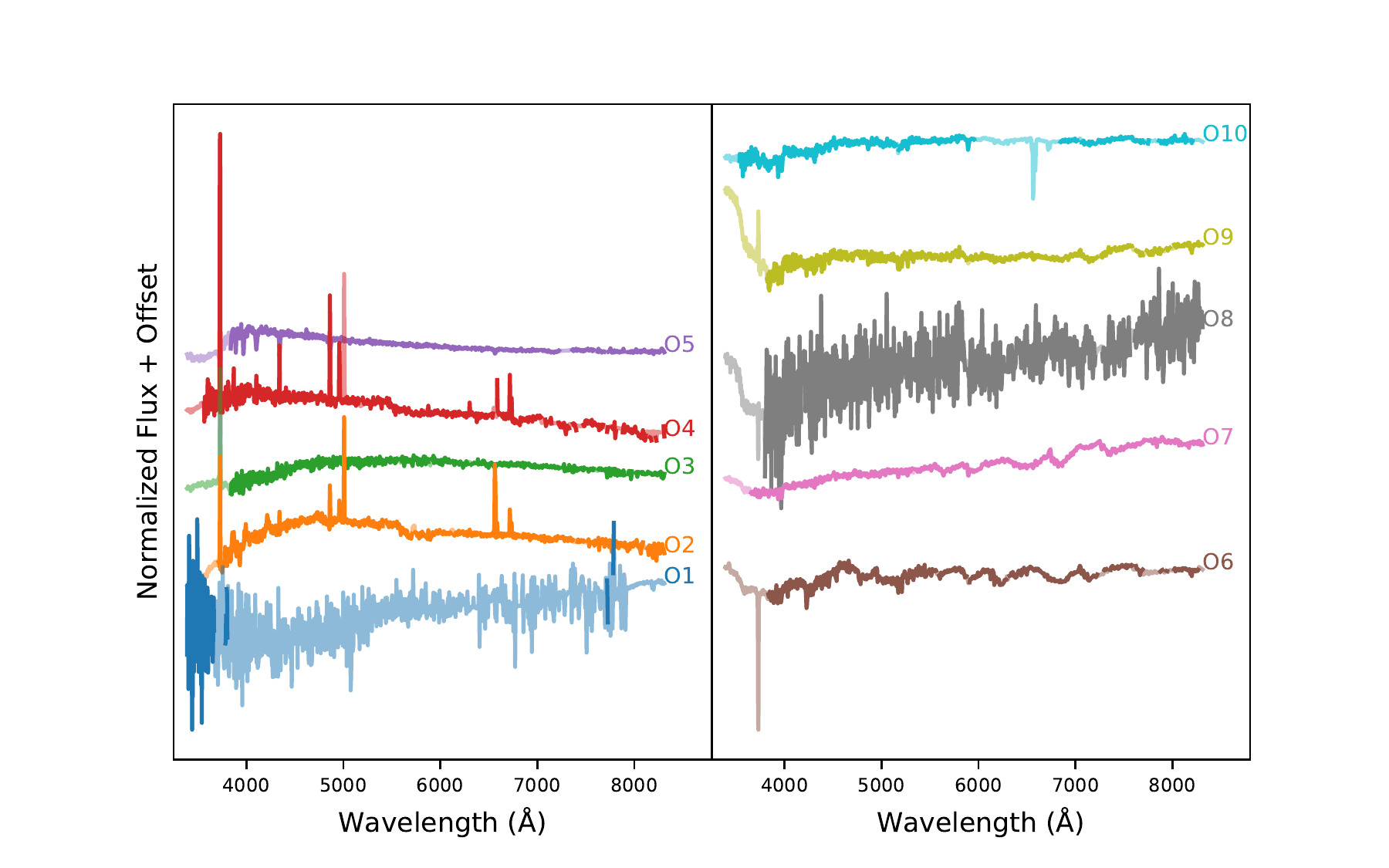}
    \caption{The top ten spectra outliers in VAE latent space using the local outlier factor algorithm, plotted with offsets added to separate the spectra. The faded parts of the spectra are bad pixels or outside the wavelength range of the original spectrum and are infilled with iterative PCA (see \autoref{sec:application}).}
    \label{fig:outliers}
\end{figure}

We consider using the VAE latent space to define outliers that are in atypical locations of latent space. These outliers could be members of rare classes or spectra that cannot be well reconstructed by the VAE. We use the local outlier factor (LOF) algorithm \citep{breunig_lof:_2000} to identify outliers. The algorithm estimates the local density of each point by using $k$ nearest neighbors and then identifies points with densities much lower than their neighbors' as outliers. The LOF algorithm is unsupervised (ie. it is not given training labels for which spectra are outliers), so we can search for outliers in the training set as well as the validation set. We use LOF with $k=20$ nearest neighbors and present the top ten outliers (labeled O1-O10 with O1 being the top outlier) in \autoref{tbl:outliers} and \autoref{fig:outliers}.

\begin{table}
\begin{center}
\begin{tabular}{ |c|c c c|c| }
 \hline
 spectrum & plate & MJD & fiber & explanation \\
 \hline
 O1 & 445 & 51873 & 68 & low SNR \\
 \hline
 O2 & 334 & 51993 & 203 & bad calibration \\
 \hline
 O3 & 480 & 51989 & 77 & close to bright star\\
 \hline
 O4 & 454 & 51908 & 607 & bad calibration \\
 \hline
 O5 & 424 & 51893 & 587 & A star\\
 \hline
 O6 & 305 & 51613 & 299 & M star\\
 \hline
 O7 & 352 & 51694 & 340 & M star \\
 \hline
 O8 & 529 & 52025 & 200 & low SNR \\
 \hline
 O9 & 276 & 51909 & 2 & M star \\
 \hline
 O10 & 414 & 51869 & 296 & missing data \\
 \hline
\end{tabular}
\end{center}
\caption{Plate-MJD-fiber IDs of the top ten outlier spectra identified in the VAE latent space, with likely explanations for why they are outlier spectra.}
\label{tbl:outliers}
\end{table}

Two of the outliers, spectra O1 and O8, are very low SNR, suggesting that the VAE is unable to map these spectra to their true latent parameters because of the severe noise. O1 is particularly severe, with most of the pixels being identified as bad pixels.

Spectrum O10 looks like a quiescent galaxy, but with a large range of data missing. The iterative PCA procedure we use to infill missing data (similar to the one used in \cite{yip_distributions_2004}) has placed strong absorption lines in the missing wavelength range. These absorption lines are unlike those seen in quiescent galaxies. Although these infilled pixels are ignored in the reconstruction loss, they are still fed into the VAE and can affect the latent parameters of the spectrum. Apparently the erroneous absorption lines had a large enough effect to make O10 an outlier in latent space.

Spectra O2 and O4 have unusual continuum shapes, with a discontinuity at around 6000 {\AA} in both cases. Taking into account the stated redshift for each spectrum, the discontinuity occurs at the wavelength separating the red and blue channels of the SDSS spectrograph. Thus, the discontinuity is likely an artifact arising from an error in calibration between the red and blue channel.

Spectrum O3 does not look like a typical galaxy spectrum, and it was taken within 10 {\arcsec} of a bright ($r = 10$) star. It is likely that this spectrum is being contaminated by stellar light, making it an outlier.

\begin{figure*}
    \plotone{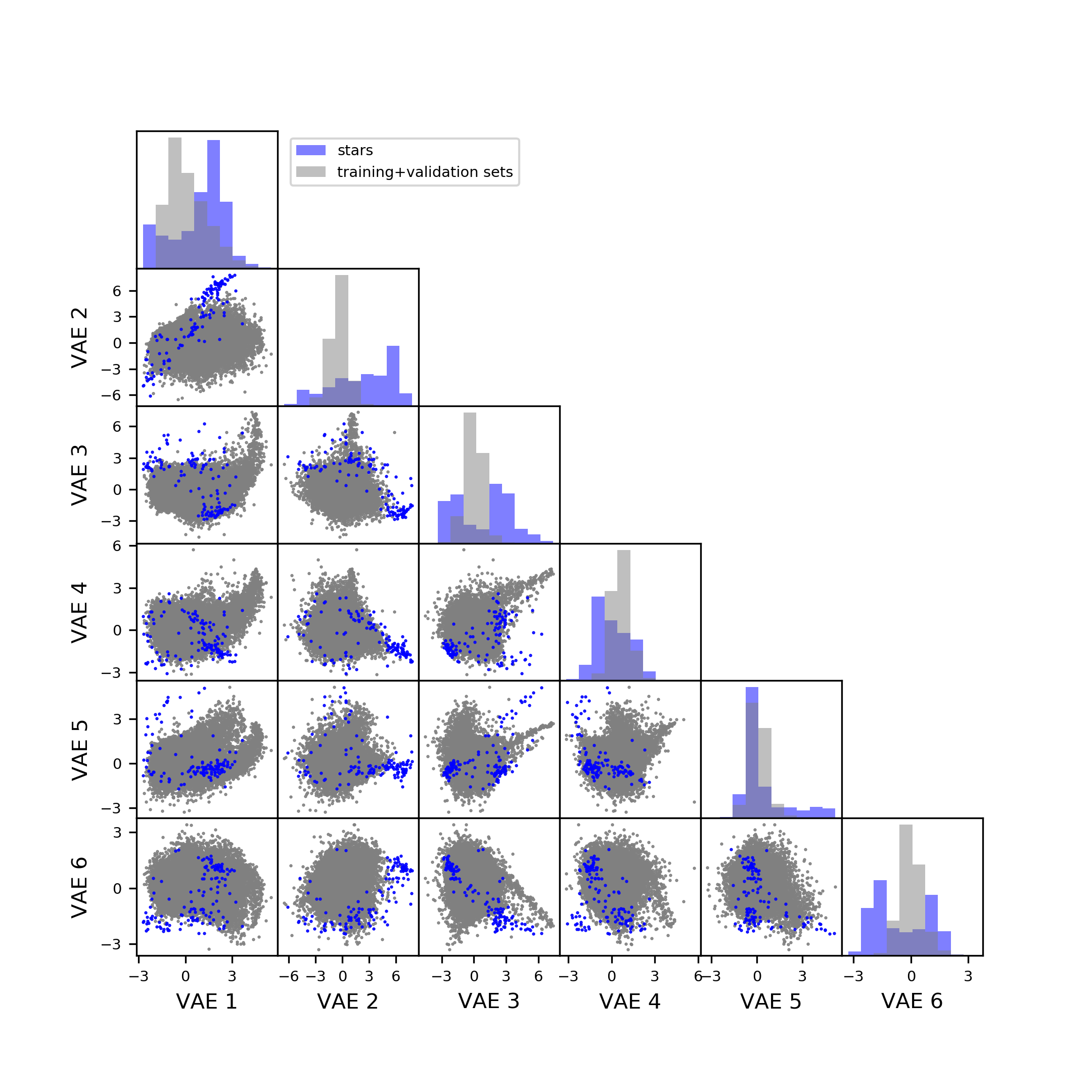}
    \caption{\edit1{Corner plot of the first six VAE components of all training and validation spectra compared with a sample of stellar spectra.}}
    \label{fig:cornerplot_stars}
\end{figure*}

Four of the outliers are stars which were erroneously classified as galaxies, suggesting that the VAE can identify these spectra as not belonging to the classes that make up the vast majority of the training set. Spectrum O5 is an A star. Spectra O6, O7, and O9 are all M stars, with O7 and O9 being spatially coincident with a galaxy. \edit1{In \autoref{fig:cornerplot_stars}, we plot the positions of a sample of SDSS stellar spectra in the VAE latent space. Some of the stellar spectra are easily identified as outliers in VAE latent space, while others overlap with the distribution of galactic spectra.}

While the top 10 outliers do not contain any new classes of objects, they are all objects that are atypical in some way, suggesting that the VAE latent space can be used to find unusual spectra. Many of the spectra are outliers because of data artifacts, but four of them are stellar spectra that somehow were classified as galaxies. Stellar spectra are rare in our set of spectra because we selected for galaxies; the fact that the VAE identifies them as outliers means that the VAE may be able to find other rare classes, given a larger dataset, or may be used to reject incorrect targeting or failed spectral extraction.

\subsection{Sensitivity to Noise}
\label{sec:noise}
\begin{figure*}
    \plottwo{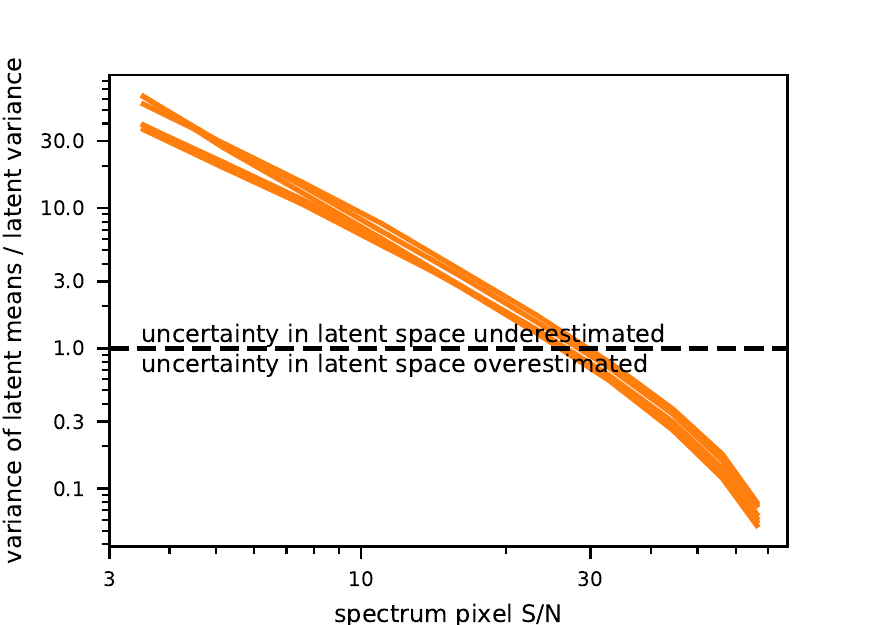}{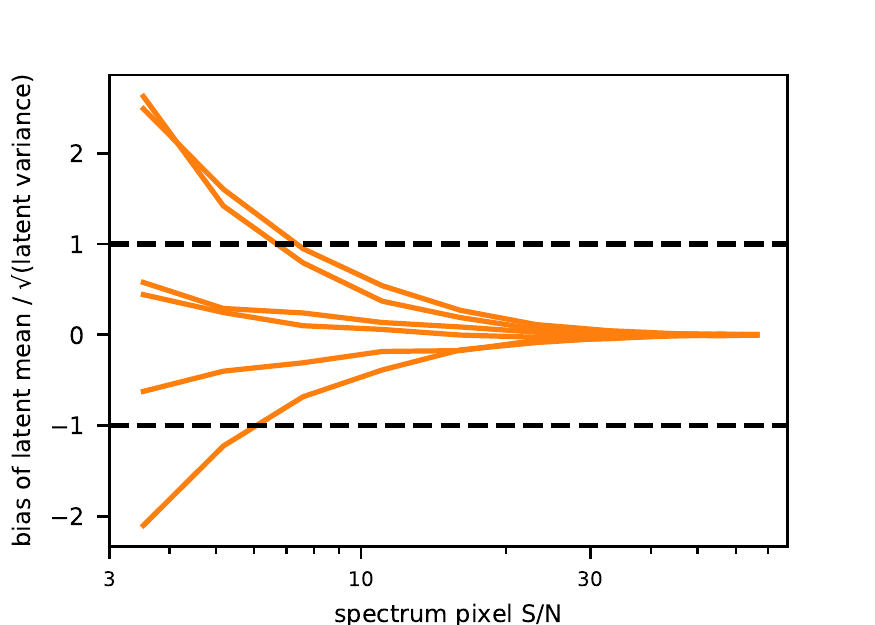}
    \caption{Variance (left) and bias (right) of the latent mean of each latent parameter (one line per parameter) when white noise is added to a high SNR spectrum, in units of the latent variance. The latent variance underestimates the uncertainty in latent space for spectra with SNR $<$ 25, which are 90\% of the spectra. The latent mean is not significantly biased if the spectrum has a SNR $\gtrapprox$ 8.}
    \label{fig:noisez}
\end{figure*}
To test the VAE's sensitivity to noise, we take the highest signal-to-noise validation spectrum (median pixel SNR $\approx$ 82) and generate many realizations of white noise. In our training set, the spectra have median pixel SNRs ranging from 5 to 88, with a median of 14. We add different levels of noise to the original spectrum to obtain simulated noisy spectra at varying SNR. Then, we encode this set of noisy spectra using the VAE, obtaining latent means and variances for each spectrum in the set. We focus on the six parameters with latent variances narrower than the prior in latent space, as these are the most important parameters for reconstruction (see \autoref{sec:recon}). For each SNR, we consider the distribution of latent means in each latent parameter obtained by encoding the noisy spectra of that SNR. These distributions look roughly Gaussian, even as strong noise is added.

The variances of these distributions of latent means is a Monte Carlo measurement of the uncertainty of a spectrum's position in latent space caused by the addition of white noise. Ideally, the latent variance returned by the VAE for a spectrum would also be an estimate of the uncertainty of that spectrum's position in latent space. We plot the ratio of these two estimates of the uncertainty, as a function of SNR, in the left panel of \autoref{fig:noisez}. The variance of the distribution of latent means increases as stronger noise is added, showing that adding noise increases the uncertainty in estimating the latent parameters of the original spectrum. By contrast, we note that the latent variances do not change drastically as noise is added, showing that the VAE does not increase its reported uncertainty in response to added noise. Thus, the variance of latent means is larger than the latent variance at low SNR and smaller at high SNR. The VAE only sees the spectrum uncertainties in training, through the definition of the reconstruction loss. That is, the VAE is not given the spectrum uncertainties at test time and thus cannot directly propagate those uncertainties to inform the latent variance. The VAE apparently does not try to estimate the level of noise from the spectrum itself to set the latent variance. Thus, the latent variance overestimates the uncertainty at high SNR and underestimates the uncertainty at low SNR. The crossing point is at SNR = 25, which is higher SNR than 90\% of the spectra, so the VAE is underestimating the uncertainty in latent space for most spectra. Adding the spectrum uncertainties as input features that the VAE sees during training could allow the VAE to better estimate the uncertainty in latent space.

The distribution of latent means would ideally be centred on the latent mean of the original spectrum, showing that the addition of noise does not bias a spectrum's position in latent space. In the right panel of \autoref{fig:noisez} we show the bias of the latent means as a function of noise level. To make the biases in different parameters comparable, we divide the bias in each parameter by the square root of the latent variance of that parameter in the original spectrum. If the biases are comparable to the latent variance, then they are problematic as they are comparable to the VAE's stated uncertainty in latent space. The biases do grow at low SNR, but remain smaller than the latent variance for SNR $>$ 8.

\section{Discussion}
\label{sec:discussion}

VAEs are able to learn an effective compression of galaxy spectra. As shown in \autoref{sec:recon}, a VAE with two parameters can reconstruct spectra as well as PCA with six components, and a VAE with six parameters can reconstruct spectra as well as PCA with ten components. Because VAEs are non-linear, they are better able to reconstruct broad spectral lines. VAEs also outperform NMF and AEs with the same number of parameters. % Unlike LLE \citep{vanderplas_reducing_2009}, reconstruction is a primary goal of VAE \authorcomment1{strike this sentence?}.

The VAE latent space is not only a good compression of spectra, but is also interpretable. The VAE can generate a synthetic spectrum for any point in latent space. In \autoref{sec:VAEtracks}, we show that traversing latent space yields continuous changes in continuum shape, line amplitude, and line width. We find tracks in the latent space that correspond to changes in star formation rate, post-starburst activity, and a transition from narrow-line to broad-line spectra. We stress that synthetic spectra are only interpretable in parts of latent space that are supported by the training set: in parts of latent space far from any training examples, the VAE is forced to extrapolate. Some of these tracks in VAE latent space have analogs in the PCA coefficient space constructed by \cite{yip_distributions_2004}. Just as we find a track from quiescent to star-forming galaxies in VAE 1, they find a similar sequence by increasing their second PCA coefficient. We also identify parts of VAE latent space occupied by extreme line-emitting galaxies and post-starburst galaxies, just as they find parts of PCA coefficient parameter space occupied by the same classes of galaxies. Because our VAE is able to handle non-linear features better than PCA, we are able to map broad-line AGN onto the same latent space as the other classes of galaxies, unlike \cite{yip_distributions_2004}.

The latent space also separates quiescent galaxies, star-forming galaxies, narrow-line AGN, and broad-line AGN, without the VAE having ever seen categorically labeled spectra. Indeed, a classification system could be made based on subdivisions of latent space. Such a system would combine the continuum sensitivity of PCA with the line amplitude sensitivity of line-ratio tests, along with line widths. Furthermore, sequences of synthetic spectra could be made that show prototypical members of each class along with transitional objects that are near class boundaries. These synthetic spectra would allow the latent space classification to be linked to other classification schemes such as PCA or line-ratio tests.

Most of the computational cost of a VAE is in training it. After training, the main cost in encoding a spectrum is de-redshifting it; encoding the de-redshifted spectrum is cheap. A VAE might even be useful in determining the redshift of a spectrum: many trial redshifts could be tested to see which one produces a de-redshifted spectrum that is best reconstructed by the VAE and thus looks similar to the training de-redshifted spectra. Generating synthetic spectra for points in latent space is also cheap. We downsample the spectra to 1000 pixels and train on only $\approx$ 47000 spectra in order to lessen the computational cost of training the VAE. Higher resolution spectra would require a larger neural net to adequately reconstruct them, which would take more epochs to train. Expanding the training set could improve VAE reconstruction or reveal interesting interpretations of latent space, but would also make each epoch of training longer.

The VAE is successful in reducing the dimensionality of spectra from 1000 pixels to the six most important VAE components; however, even this smaller, six dimensional space is difficult to fully explore. While highly informative, the set of scatter plots in \autoref{fig:cornerplot} are 2D projections that do not capture the full structure of latent space. Just as we use the \texttt{spectro1d} and line-ratio classifications to interpret the latent space, other classifications could be used to create further interpretations. Simulated spectra could also be given to the VAE to decode to see how the simulated galaxies' properties map onto latent space. Clustering techniques like k-means or DBSCAN \citep{ester_density-based_1996} could be used to find meaningful divisions of latent space in an unsupervised manner. These clustering techniques should be faster to run and perhaps more effective in the smaller latent space, rather than the original space of spectra. Interactive visualizations could also be useful in allowing astronomers to explore latent space. With an interactive visualization, different latent parameters could be scatterplotted, and the user could quickly look at the spectra that correspond to different points in latent space or color the scatterplot by some measured property of the spectra. \cite{2019arXiv191106823R} reduce the dimensionality of the SDSS spectra with Uniform Manifold Approximation and Projection and present an interactive visualization\footnote{Available at \url{https://galaxyportal.space/}}.

We make a first effort to find outlier spectra in \autoref{sec:outliers}, but a search for rare spectra would need to involve a much larger dataset. The VAE could be trained on the entire SDSS main galaxy sample, for example, to then find the most unusual spectra. Techniques like DEMUD \citep{wagstaff_guiding_2013} would be useful in finding interesting spectra. Again, the reduced dimensionality of the latent space should make outlier searches easier.

The VAE is able to encode spectra with missing data and levels of noise, but uncertainties in the data could be handled better. In our VAE, both missing data and measurement uncertainty are handled through weighting the data pixels in the reconstruction loss: good pixels are inverse-variance weighted and bad pixels are given a weight of zero. We find that setting a noise floor (ie. a maximum weight) improves the stability of training the VAE. Excluding the bad pixels from the reconstruction loss means that the VAE is unconcerned with reconstructing the fluxes in these pixels. However, the fluxes in these bad pixels are still propagated through the encoder, possibly affecting the latent representation. We find that some sort of imputation is necessary for the VAE to handle bad pixels well, and we use the iterative PCA procedure presented in \cite{yip_distributions_2004}. One could imagine an iterative VAE procedure where bad pixels are imputed with this VAE in order to train a second VAE. In \autoref{sec:noise}, we find that degrading spectra down to a pixel SNR of 8 does not drastically affect the latent means. However, we find that the latent variance does not grow when the SNR is degraded, meaning that the VAE is reporting the same variance in latent space regardless of SNR. This behavior is undesired because the VAE is underpredicting the uncertainty in latent space for low SNR spectra. The pixel-level noise only appears in the reconstruction loss and is not included as a feature that the VAE can learn the variance from. Including the pixel-level noise as a feature may allow the VAE to learn appropriate latent variances for low SNR spectra. 

Just as can be done in PCA and LLE, the VAE can be extended to include measurements beyond spectra, eg. fluxes or shape parameters. Some procedure to normalize the data is necessary for any dimensionality reduction technique. Treating the reconstruction loss as a negative log likelihood offers a natural prescription: additional measurements should be added to the loss function with inverse-variance weights. 

Augmenting the VAE using multi-task learning \citep{caruana_multitask_1993} could improve the quality of the VAE latent representions as well as the accuracy of classification or regression tasks on the input spectra. In multi-task learning, one neural network is trained to perform multiple related tasks, which often yields improvements compared to training a separate network for each task. Since a set of neurons is being used for all tasks, this set of neurons tends to learn a representation that generalizes well across tasks, avoiding overfitting on any one task. Our VAE could be augmented into a semi-supervised autoencoder \citep{haiyan_semi-supervised_2015,zhuang_representation_2015} that also performs an additional task like classification or regression (eg. on star formation rate). The network that performs the additional task could take the encoder's output as its input, sharing no weights with the decoder and producing its own output. We note that training the VAE and then training a network for the additional task that uses the VAE latent means as features is not multi-task learning. Instead, training the encoder, decoder, and the additional task simultaneously (eg. alternating tasks every epoch) would constitute multi-task learning.

\section{Conclusion}
\label{sec:conclusion}

Efficiently exploring and classifying large astronomical datasets is an important problem that can be addressed with dimensionality reduction techniques. In this work, we have demonstrated that a VAE be used to reduce the dimensionality of SDSS spectra to six latent parameters while retaining enough information to accurately reconstruct individual spectra. Due to its non-linear behavior, the VAE can capture non-linear features, such as line widths, with fewer parameters than PCA. Unlike line-ratio diagnostics, the VAE also uses the continuum information in the spectrum. The VAE latent space is interpretable and shows clear separations between different classes of galaxies, even though the VAE was never given these classifications in training. Tracks in latent space yield sequences of spectra whose physical properties smoothly vary, and unusual objects can be identified as outliers within the latent space. While even a six parameter latent space is difficult to fully visualize and interpret, reducing the dimensionality of the spectra makes them more amenable to both computational techniques and human scrutiny. This latent space can be more fully explored using, for example, clustering techniques, outlier searches, and interactive visualizations. VAEs are a fast dimensionality reduction technique that yields compact, interpretable latent spaces and have the potential to enable the fast exploration and classification of large astronomical datasets.

\acknowledgments

S.K.N.P. acknowledges support from the DIRAC Institute in the Department of Astronomy at the University of Washington. The DIRAC Institute is supported through generous gifts from the Charles and Lisa Simonyi Fund for Arts and Sciences, and the Washington Research Foundation. J.R.V. acknowledges support from FONDECYT postdoctoral grants 3160772 and the Ministry of Economy, Development, and Tourism's Millennium Science Initiative through grant IC12009, awarded to The Millennium Institute of Astrophysics, MAS. The authors also acknowledge partial support from NSF grants AST-1715122 and OAC-1739419.

%This is the DR7 acknowledgement
Funding for the SDSS and SDSS-II has been provided by the Alfred P. Sloan Foundation, the Participating Institutions, the National Science Foundation, the U.S. Department of Energy, the National Aeronautics and Space Administration, the Japanese Monbukagakusho, the Max Planck Society, and the Higher Education Funding Council for England. The SDSS Web Site is \url{http://www.sdss.org/}.

The SDSS is managed by the Astrophysical Research Consortium for the Participating Institutions. The Participating Institutions are the American Museum of Natural History, Astrophysical Institute Potsdam, University of Basel, University of Cambridge, Case Western Reserve University, University of Chicago, Drexel University, Fermilab, the Institute for Advanced Study, the Japan Participation Group, Johns Hopkins University, the Joint Institute for Nuclear Astrophysics, the Kavli Institute for Particle Astrophysics and Cosmology, the Korean Scientist Group, the Chinese Academy of Sciences (LAMOST), Los Alamos National Laboratory, the Max-Planck-Institute for Astronomy (MPIA), the Max-Planck-Institute for Astrophysics (MPA), New Mexico State University, Ohio State University, University of Pittsburgh, University of Portsmouth, Princeton University, the United States Naval Observatory, and the University of Washington.


\vspace{5mm}
\facilities{Sloan}


\software{astropy \citep{the_astropy_collaboration_astropy:_2013,the_astropy_collaboration_astropy_2018},
scikit-learn \citep{pedregosa_scikit-learn:_2011},
AstroML \citep{vanderplas_introduction_2012},
pytorch \citep{paszke_automatic_2017}}

\bibliography{references}

\begin{thebibliography}{}
\expandafter\ifx\csname natexlab\endcsname\relax\def\natexlab#1{#1}\fi
\providecommand{\url}[1]{\href{#1}{#1}}

\bibitem[{Abazajian {et~al.}(2009)Abazajian, Adelman-McCarthy, Agüeros, Allam,
  Prieto, An, Anderson, Anderson, Annis, Bahcall, Bailer-Jones, Barentine,
  Bassett, Becker, Beers, Bell, Belokurov, Berlind, Berman, Bernardi,
  Bickerton, Bizyaev, Blakeslee, Blanton, Bochanski, Boroski, Brewington,
  Brinchmann, Brinkmann, Brunner, Budavári, Carey, Carliles, Carr, Castander,
  Cinabro, Connolly, Csabai, Cunha, Czarapata, Davenport, de~Haas, Dilday, Doi,
  Eisenstein, Evans, Evans, Fan, Friedman, Frieman, Fukugita, Gänsicke, Gates,
  Gillespie, Gilmore, Gonzalez, Gonzalez, Grebel, Gunn, Györy, Hall, Harding,
  Harris, Harvanek, Hawley, Hayes, Heckman, Hendry, Hennessy, Hindsley,
  Hoblitt, Hogan, Hogg, Holtzman, Hyde, Ichikawa, Ichikawa, Im, Ivezić,
  Jester, Jiang, Johnson, Jorgensen, Jurić, Kent, Kessler, Kleinman, Knapp,
  Konishi, Kron, Krzesinski, Kuropatkin, Lampeitl, Lebedeva, Lee, Lee, Leger,
  Lépine, Li, Lima, Lin, Long, Loomis, Loveday, Lupton, Magnier, Malanushenko,
  Malanushenko, Mandelbaum, Margon, Marriner, Martínez-Delgado, Matsubara,
  McGehee, McKay, Meiksin, Morrison, Mullally, Munn, Murphy, Nash, Nebot,
  Neilsen, Newberg, Newman, Nichol, Nicinski, Nieto-Santisteban, Nitta,
  Okamura, Oravetz, Ostriker, Owen, Padmanabhan, Pan, Park, Pauls, Peoples,
  Percival, Pier, Pope, Pourbaix, Price, Purger, Quinn, Raddick, Fiorentin,
  Richards, Richmond, Riess, Rix, Rockosi, Sako, Schlegel, Schneider, Scholz,
  Schreiber, Schwope, Seljak, Sesar, Sheldon, Shimasaku, Sibley, Simmons,
  Sivarani, Smith, Smith, Smolčić, Snedden, Stebbins, Steinmetz, Stoughton,
  Strauss, SubbaRao, Suto, Szalay, Szapudi, Szkody, Tanaka, Tegmark, Teodoro,
  Thakar, Tremonti, Tucker, Uomoto, Vanden~Berk, Vandenberg, Vidrih, Vogeley,
  Voges, Vogt, Wadadekar, Watters, Weinberg, West, White, Wilhite, Wonders,
  Yanny, Yocum, York, Zehavi, Zibetti, \& Zucker}]{abazajian_seventh_2009}
Abazajian, K.~N., Adelman-McCarthy, J.~K., Agüeros, M.~A., {et~al.} 2009, The
  Astrophysical Journal Supplement Series, 182, 543.
\newblock
  \url{http://stacks.iop.org/0067-0049/182/i=2/a=543?key=crossref.bc07496fb06b943bcf82755687fa84b4}

\bibitem[{Almeida {et~al.}(2010)Almeida, Aguerri, Muñoz-Tuñón, \&
  de~Vicente}]{almeida_automatic_2010}
Almeida, J.~S., Aguerri, J. A.~L., Muñoz-Tuñón, C., \& de~Vicente, A. 2010,
  The Astrophysical Journal, 714, 487.
\newblock
  \url{http://stacks.iop.org/0004-637X/714/i=1/a=487?key=crossref.ce6d36845688e0baef3e542764ff5365}

\bibitem[{Baldwin {et~al.}(1981)Baldwin, Phillips, \&
  Terlevich}]{baldwin_classification_1981}
Baldwin, J.~A., Phillips, M.~M., \& Terlevich, R. 1981, Publications of the
  Astronomical Society of the Pacific, 93, 5.
\newblock \url{http://iopscience.iop.org/article/10.1086/130766}

\bibitem[{Ball {et~al.}(2004)Ball, Loveday, Fukugita, Nakamura, Okamura,
  Brinkmann, \& Brunner}]{ball_galaxy_2004}
Ball, N.~M., Loveday, J., Fukugita, M., {et~al.} 2004, Monthly Notices of the
  Royal Astronomical Society, 348, 1038.
\newblock
  \url{https://academic.oup.com/mnras/article-lookup/doi/10.1111/j.1365-2966.2004.07429.x}

\bibitem[{{Baron} \& {Poznanski}(2017)}]{2017MNRAS.465.4530B}
{Baron}, D., \& {Poznanski}, D. 2017, \mnras, 465, 4530

\bibitem[{Breunig {et~al.}(2000)Breunig, Kriegel, Ng, \&
  Sander}]{breunig_lof:_2000}
Breunig, M.~M., Kriegel, H.-P., Ng, R.~T., \& Sander, J. 2000, ACM SIGMOD
  Record, 29, 93.
\newblock \url{http://portal.acm.org/citation.cfm?doid=335191.335388}

\bibitem[{Caruana(1993)}]{caruana_multitask_1993}
Caruana, R.~A. 1993, in Machine {Learning} {Proceedings} 1993 (Elsevier),
  41--48.
\newblock
  \url{https://linkinghub.elsevier.com/retrieve/pii/B9781558603073500125}

\bibitem[{Chardin {et~al.}(2019)Chardin, Uhlrich, Aubert, Deparis, Gillet,
  Ocvirk, \& Lewis}]{chardin_deep_2019}
Chardin, J., Uhlrich, G., Aubert, D., {et~al.} 2019, Monthly Notices of the
  Royal Astronomical Society, 490, 1055.
\newblock \url{https://academic.oup.com/mnras/article/490/1/1055/5584889}

\bibitem[{Ester {et~al.}(1996)Ester, Kriegel, Sander, \&
  Xu}]{ester_density-based_1996}
Ester, M., Kriegel, H.-P., Sander, J., \& Xu, X. 1996, in Proceedings of 2nd
  {International} {Conference} on {Knowledge} {Discovery} and {Data} {Mining},
  226--231

\bibitem[{Folkes {et~al.}(1996)Folkes, Lahav, \&
  Maddox}]{folkes_artificial_1996}
Folkes, S.~R., Lahav, O., \& Maddox, S.~J. 1996, Monthly Notices of the Royal
  Astronomical Society, 283, 651.
\newblock
  \url{https://academic.oup.com/mnras/article-lookup/doi/10.1093/mnras/283.2.651}

\bibitem[{Goodfellow {et~al.}(2016)Goodfellow, Bengio, \&
  Courville}]{Goodfellow-et-al-2016}
Goodfellow, I., Bengio, Y., \& Courville, A. 2016, Deep Learning (MIT Press),
  \url{http://www.deeplearningbook.org}

\bibitem[{Gretton {et~al.}(2007)Gretton, Borgwardt, Rasch, Sch\"{o}lkopf, \&
  Smola}]{NIPS2006_3110}
Gretton, A., Borgwardt, K., Rasch, M., Sch\"{o}lkopf, B., \& Smola, A.~J. 2007,
  in Advances in Neural Information Processing Systems 19, ed.
  B.~Sch\"{o}lkopf, J.~C. Platt, \& T.~Hoffman (MIT Press), 513--520.
\newblock
  \url{http://papers.nips.cc/paper/3110-a-kernel-method-for-the-two-sample-problem.pdf}

\bibitem[{Gunn {et~al.}(2006)Gunn, Siegmund, Mannery, Owen, Hull, Leger, Carey,
  Knapp, York, Boroski, Kent, Lupton, Rockosi, Evans, Waddell, Anderson, Annis,
  Barentine, Bartoszek, Bastian, Bracker, Brewington, Briegel, Brinkmann,
  Brown, Carr, Czarapata, Drennan, Dombeck, Federwitz, Gillespie, Gonzales,
  Hansen, Harvanek, Hayes, Jordan, Kinney, Klaene, Kleinman, Kron, Kresinski,
  Lee, Limmongkol, Lindenmeyer, Long, Loomis, McGehee, Mantsch, Neilsen,
  Neswold, Newman, Nitta, Peoples, Pier, Prieto, Prosapio, Rivetta, Schneider,
  Snedden, \& Wang}]{gunn_2.5_2006}
Gunn, J.~E., Siegmund, W.~A., Mannery, E.~J., {et~al.} 2006, The Astronomical
  Journal, 131, 2332.
\newblock \url{http://stacks.iop.org/1538-3881/131/i=4/a=2332}

\bibitem[{Haiyan {et~al.}(2015)Haiyan, Haomin, Xueming, \&
  Haijun}]{haiyan_semi-supervised_2015}
Haiyan, W., Haomin, Y., Xueming, L., \& Haijun, R. 2015, in 2015
  {International} {Conference} on {Computational} {Intelligence} and
  {Communication} {Networks} ({CICN}) (Jabalpur, India: IEEE), 1424--1430.
\newblock \url{http://ieeexplore.ieee.org/document/7546333/}

\bibitem[{Higgins {et~al.}(2017)Higgins, Matthey, Pal, Burgess, Glorot,
  Botvinick, Mohamed, \& Lerchner}]{higgins_-vae:_2017}
Higgins, I., Matthey, L., Pal, A., {et~al.} 2017, International Conference on
  Learning Representations, 22

\bibitem[{Iwasaki {et~al.}(2019)Iwasaki, Ichinohe, \&
  Uchiyama}]{iwasaki_x-ray_2019}
Iwasaki, H., Ichinohe, Y., \& Uchiyama, Y. 2019, Monthly Notices of the Royal
  Astronomical Society, 488, 4106.
\newblock \url{https://academic.oup.com/mnras/article/488/3/4106/5538838}

\bibitem[{Kewley {et~al.}(2001)Kewley, Dopita, Sutherland, Heisler, \&
  Trevena}]{kewley_theoretical_2001}
Kewley, L.~J., Dopita, M.~A., Sutherland, R.~S., Heisler, C.~A., \& Trevena, J.
  2001, The Astrophysical Journal, 556, 121.
\newblock \url{http://stacks.iop.org/0004-637X/556/i=1/a=121}

\bibitem[{Kewley {et~al.}(2006)Kewley, Groves, Kauffmann, \&
  Heckman}]{kewley_host_2006}
Kewley, L.~J., Groves, B., Kauffmann, G., \& Heckman, T. 2006, Monthly Notices
  of the Royal Astronomical Society, 372, 961.
\newblock
  \url{https://academic.oup.com/mnras/article-lookup/doi/10.1111/j.1365-2966.2006.10859.x}

\bibitem[{Kewley {et~al.}(2019)Kewley, Nicholls, \&
  Sutherland}]{kewley_understanding_2019}
Kewley, L.~J., Nicholls, D.~C., \& Sutherland, R.~S. 2019, Annual Review of
  Astronomy and Astrophysics, 57, 511.
\newblock
  \url{https://www.annualreviews.org/doi/10.1146/annurev-astro-081817-051832}

\bibitem[{Kingma \& Ba(2015)}]{kingma_adam:_2015}
Kingma, D.~P., \& Ba, J. 2015, International Conference on Learning
  Representations, arXiv: 1412.6980.
\newblock \url{http://arxiv.org/abs/1412.6980}

\bibitem[{Kingma \& Welling(2013)}]{kingma_auto-encoding_2013}
Kingma, D.~P., \& Welling, M. 2013, International Conference on Learning
  Representations, arXiv: 1312.6114.
\newblock \url{http://arxiv.org/abs/1312.6114}

\bibitem[{Kullback \& Leibler(1951)}]{kullback1951}
Kullback, S., \& Leibler, R.~A. 1951, Ann. Math. Statist., 22, 79.
\newblock \url{https://doi.org/10.1214/aoms/1177729694}

\bibitem[{Lawlor {et~al.}(2016)Lawlor, Budavári, \&
  Mahoney}]{lawlor_mapping_2016}
Lawlor, D., Budavári, T., \& Mahoney, M.~W. 2016, The Astrophysical Journal,
  833, 26.
\newblock
  \url{http://stacks.iop.org/0004-637X/833/i=1/a=26?key=crossref.1973aaef4535b89994cb47aec470bcfe}

\bibitem[{Li {et~al.}(2017)Li, Pan, \& Duan}]{li_parameterizing_2017}
Li, X.-R., Pan, R.-Y., \& Duan, F.-Q. 2017, Research in Astronomy and
  Astrophysics, 17, 036.
\newblock
  \url{http://stacks.iop.org/1674-4527/17/i=4/a=036?key=crossref.37316d277372e3cf074c88a37439df64}

\bibitem[{Lu {et~al.}(2006)Lu, Zhou, Wang, Wang, Dong, Zhuang, \&
  Li}]{lu_ensemble_2006}
Lu, H., Zhou, H., Wang, J., {et~al.} 2006, The Astronomical Journal, 131, 790.
\newblock \url{http://stacks.iop.org/1538-3881/131/i=2/a=790}

\bibitem[{Ma {et~al.}(2019)Ma, Xu, Zhu, Hu, Li, Shan, Zhu, Gu, Li, Liu, \&
  Wu}]{ma_machine_2019}
Ma, Z., Xu, H., Zhu, J., {et~al.} 2019, The Astrophysical Journal Supplement
  Series, 240, 34.
\newblock
  \url{http://stacks.iop.org/0067-0049/240/i=2/a=34?key=crossref.844310e65f333210dddd2f7323c6ab2e}

\bibitem[{Meusinger {et~al.}(2017)Meusinger, Brünecke, Schalldach, \& in~der
  Au}]{meusinger_large_2017}
Meusinger, H., Brünecke, J., Schalldach, P., \& in~der Au, A. 2017, Astronomy
  \& Astrophysics, 597, A134.
\newblock \url{http://www.aanda.org/10.1051/0004-6361/201629139}

\bibitem[{Meusinger {et~al.}(2012)Meusinger, Schalldach, Scholz, in~der Au,
  Newholm, de~Hoon, \& Kaminsky}]{meusinger_unusual_2012}
Meusinger, H., Schalldach, P., Scholz, R.-D., {et~al.} 2012, Astronomy \&
  Astrophysics, 541, A77.
\newblock \url{http://www.aanda.org/10.1051/0004-6361/201118143}

\bibitem[{Naul {et~al.}(2018)Naul, Bloom, PÃ©rez, \& van~der
  Walt}]{naul_recurrent_2018}
Naul, B., Bloom, J.~S., PÃ©rez, F., \& van~der Walt, S. 2018, Nature
  Astronomy, 2, 151.
\newblock \url{http://www.nature.com/articles/s41550-017-0321-z}

\bibitem[{Osterbrock \& de~Robertis(1985)}]{osterbrock_optical_1985}
Osterbrock, D.~E., \& de~Robertis, M.~M. 1985, Publications of the Astronomical
  Society of the Pacific, 97, 1129.
\newblock \url{http://iopscience.iop.org/article/10.1086/131676}

\bibitem[{Paszke {et~al.}(2017)Paszke, Gross, Chintala, Chanan, Yang, DeVito,
  Lin, Desmaison, Antiga, \& Lerer}]{paszke_automatic_2017}
Paszke, A., Gross, S., Chintala, S., {et~al.} 2017, Conference on Neural
  Information Processing Systems, 4

\bibitem[{Pedregosa {et~al.}(2011)Pedregosa, Varoquaux, Gramfort, Michel,
  Thirion, Grisel, Blondel, Prettenhofer, Weiss, Dubourg, Vanderplas, Passos,
  \& Cournapeau}]{pedregosa_scikit-learn:_2011}
Pedregosa, F., Varoquaux, G., Gramfort, A., {et~al.} 2011, Journal of Machine
  Learning Research, 12, 2825

\bibitem[{{Reis} {et~al.}(2018){Reis}, {Poznanski}, {Baron}, {Zasowski}, \&
  {Shahaf}}]{2018MNRAS.476.2117R}
{Reis}, I., {Poznanski}, D., {Baron}, D., {Zasowski}, G., \& {Shahaf}, S. 2018,
  \mnras, 476, 2117

\bibitem[{{Reis} {et~al.}(2019){Reis}, {Rotman}, {Poznanski}, {Prochaska}, \&
  {Wolf}}]{2019arXiv191106823R}
{Reis}, I., {Rotman}, M., {Poznanski}, D., {Prochaska}, J.~X., \& {Wolf}, L.
  2019, arXiv e-prints, arXiv:1911.06823

\bibitem[{Richards {et~al.}(2009)Richards, Freeman, Lee, \&
  Schafer}]{richards_exploiting_2009}
Richards, J.~W., Freeman, P.~E., Lee, A.~B., \& Schafer, C.~M. 2009, The
  Astrophysical Journal, 691, 32.
\newblock
  \url{http://stacks.iop.org/0004-637X/691/i=1/a=32?key=crossref.a4e31a29387a5c286ed9ece22bdf19f7}

\bibitem[{Smee {et~al.}(2013)Smee, Gunn, Uomoto, Roe, Schlegel, Rockosi, Carr,
  Leger, Dawson, Olmstead, Brinkmann, Owen, Barkhouser, Honscheid, Harding,
  Long, Lupton, Loomis, Anderson, Annis, Bernardi, Bhardwaj, Bizyaev, Bolton,
  Brewington, Briggs, Burles, Burns, Castander, Connolly, Davenport, Ebelke,
  Epps, Feldman, Friedman, Frieman, Heckman, Hull, Knapp, Lawrence, Loveday,
  Mannery, Malanushenko, Malanushenko, Merrelli, Muna, Newman, Nichol, Oravetz,
  Pan, Pope, Ricketts, Shelden, Sandford, Siegmund, Simmons, Smith, Snedden,
  Schneider, SubbaRao, Tremonti, Waddell, \& York}]{smee_multi-object_2013}
Smee, S.~A., Gunn, J.~E., Uomoto, A., {et~al.} 2013, The Astronomical Journal,
  146, 32.
\newblock
  \url{http://stacks.iop.org/1538-3881/146/i=2/a=32?key=crossref.d156471580f8c4459650597033951b06}

\bibitem[{Stoughton {et~al.}(2002)Stoughton, Lupton, Bernardi, Blanton, Burles,
  Castander, Connolly, Eisenstein, Frieman, Hennessy, Hindsley, Ivezić, Kent,
  Kunszt, Lee, Meiksin, Munn, Newberg, Nichol, Nicinski, Pier, Richards,
  Richmond, Schlegel, Smith, Strauss, SubbaRao, Szalay, Thakar, Tucker,
  Vanden~Berk, Yanny, Adelman, Anderson, Anderson, Annis, Bahcall, Bakken,
  Bartelmann, Bastian, Bauer, Berman, Böhringer, Boroski, Bracker, Briegel,
  Briggs, Brinkmann, Brunner, Carey, Carr, Chen, Christian, Colestock, Crocker,
  Csabai, Czarapata, Dalcanton, Davidsen, Davis, Dehnen, Dodelson, Doi,
  Dombeck, Donahue, Ellman, Elms, Evans, Eyer, Fan, Federwitz, Friedman,
  Fukugita, Gal, Gillespie, Glazebrook, Gray, Grebel, Greenawalt, Greene, Gunn,
  de~Haas, Haiman, Haldeman, Hall, Hamabe, Hansen, Harris, Harris, Harvanek,
  Hawley, Hayes, Heckman, Helmi, Henden, Hogan, Hogg, Holmgren, Holtzman,
  Huang, Hull, Ichikawa, Ichikawa, Johnston, Kauffmann, Kim, Kimball, Kinney,
  Klaene, Kleinman, Klypin, Knapp, Korienek, Krolik, Kron, Krzesiński, Lamb,
  Leger, Limmongkol, Lindenmeyer, Long, Loomis, Loveday, MacKinnon, Mannery,
  Mantsch, Margon, McGehee, McKay, McLean, Menou, Merelli, Mo, Monet, Nakamura,
  Narayanan, Nash, Neilsen, Newman, Nitta, Odenkirchen, Okada, Okamura,
  Ostriker, Owen, Pauls, Peoples, Peterson, Petravick, Pope, Pordes, Postman,
  Prosapio, Quinn, Rechenmacher, Rivetta, Rix, Rockosi, Rosner, Ruthmansdorfer,
  Sandford, Schneider, Scranton, Sekiguchi, Sergey, Sheth, Shimasaku, Smee,
  Snedden, Stebbins, Stubbs, Szapudi, Szkody, Szokoly, Tabachnik, Tsvetanov,
  Uomoto, Vogeley, Voges, Waddell, Walterbos, Wang, Watanabe, Weinberg, White,
  White, Wilhite, Wolfe, Yasuda, York, Zehavi, \& Zheng}]{stoughton_sloan_2002}
Stoughton, C., Lupton, R.~H., Bernardi, M., {et~al.} 2002, The Astronomical
  Journal, 123, 485.
\newblock \url{http://stacks.iop.org/1538-3881/123/i=1/a=485}

\bibitem[{Strauss {et~al.}(2002)Strauss, Weinberg, Lupton, Narayanan, Annis,
  Bernardi, Blanton, Burles, Connolly, Dalcanton, Doi, Eisenstein, Frieman,
  Fukugita, Gunn, Ivezić, Kent, Kim, Knapp, Kron, Munn, Newberg, Nichol,
  Okamura, Quinn, Richmond, Schlegel, Shimasaku, SubbaRao, Szalay, Vanden~Berk,
  Vogeley, Yanny, Yasuda, York, \& Zehavi}]{strauss_spectroscopic_2002}
Strauss, M.~A., Weinberg, D.~H., Lupton, R.~H., {et~al.} 2002, The Astronomical
  Journal, 124, 1810.
\newblock \url{http://stacks.iop.org/1538-3881/124/i=3/a=1810}

\bibitem[{{The Astropy Collaboration} {et~al.}(2013){The Astropy
  Collaboration}, Robitaille, Tollerud, Greenfield, Droettboom, Bray, Aldcroft,
  Davis, Ginsburg, Price-Whelan, Kerzendorf, Conley, Crighton, Barbary, Muna,
  Ferguson, Grollier, Parikh, Nair, Günther, Deil, Woillez, Conseil, Kramer,
  Turner, Singer, Fox, Weaver, Zabalza, Edwards, Azalee~Bostroem, Burke, Casey,
  Crawford, Dencheva, Ely, Jenness, Labrie, Lim, Pierfederici, Pontzen, Ptak,
  Refsdal, Servillat, \& Streicher}]{the_astropy_collaboration_astropy:_2013}
{The Astropy Collaboration}, Robitaille, T.~P., Tollerud, E.~J., {et~al.} 2013,
  Astronomy \& Astrophysics, 558, A33.
\newblock \url{http://www.aanda.org/10.1051/0004-6361/201322068}

\bibitem[{{The Astropy Collaboration} {et~al.}(2018){The Astropy
  Collaboration}, Price-Whelan, Sipőcz, Günther, Lim, Crawford, Conseil,
  Shupe, Craig, Dencheva, Ginsburg, VanderPlas, Bradley, Pérez-Suárez,
  de~Val-Borro, {(Primary Paper Contributors)}, Aldcroft, Cruz, Robitaille,
  Tollerud, {(Astropy Coordination Committee)}, Ardelean, Babej, Bach,
  Bachetti, Bakanov, Bamford, Barentsen, Barmby, Baumbach, Berry, Biscani,
  Boquien, Bostroem, Bouma, Brammer, Bray, Breytenbach, Buddelmeijer, Burke,
  Calderone, Rodríguez, Cara, Cardoso, Cheedella, Copin, Corrales, Crichton,
  D’Avella, Deil, Depagne, Dietrich, Donath, Droettboom, Earl, Erben, Fabbro,
  Ferreira, Finethy, Fox, Garrison, Gibbons, Goldstein, Gommers, Greco,
  Greenfield, Groener, Grollier, Hagen, Hirst, Homeier, Horton, Hosseinzadeh,
  Hu, Hunkeler, Ivezić, Jain, Jenness, Kanarek, Kendrew, Kern, Kerzendorf,
  Khvalko, King, Kirkby, Kulkarni, Kumar, Lee, Lenz, Littlefair, Ma, Macleod,
  Mastropietro, McCully, Montagnac, Morris, Mueller, Mumford, Muna, Murphy,
  Nelson, Nguyen, Ninan, Nöthe, Ogaz, Oh, Parejko, Parley, Pascual, Patil,
  Patil, Plunkett, Prochaska, Rastogi, Janga, Sabater, Sakurikar, Seifert,
  Sherbert, Sherwood-Taylor, Shih, Sick, Silbiger, Singanamalla, Singer,
  Sladen, Sooley, Sornarajah, Streicher, Teuben, Thomas, Tremblay, Turner,
  Terrón, Kerkwijk, de~la Vega, Watkins, Weaver, Whitmore, Woillez, Zabalza,
  \& {(Astropy Contributors)}}]{the_astropy_collaboration_astropy_2018}
{The Astropy Collaboration}, Price-Whelan, A.~M., Sipőcz, B.~M., {et~al.}
  2018, The Astronomical Journal, 156, 123.
\newblock \url{https://iopscience.iop.org/article/10.3847/1538-3881/aabc4f}

\bibitem[{Tr\"{o}ster {et~al.}(2019)Tr\"{o}ster, Ferguson, Harnois-D\'{e}raps,
  \& McCarthy}]{troster_painting_2019}
Tr\"{o}ster, T., Ferguson, C., Harnois-D\'{e}raps, J., \& McCarthy, I.~G. 2019,
  Monthly Notices of the Royal Astronomical Society: Letters, 487, L24.
\newblock \url{https://academic.oup.com/mnrasl/article/487/1/L24/5505854}

\bibitem[{Tsang \& Schultz(2019)}]{tsang_deep_2019}
Tsang, B. T.-H., \& Schultz, W.~C. 2019, The Astrophysical Journal, 877, L14.
\newblock \url{https://iopscience.iop.org/article/10.3847/2041-8213/ab212c}

\bibitem[{Vanderplas \& Connolly(2009)}]{vanderplas_reducing_2009}
Vanderplas, J., \& Connolly, A. 2009, The Astronomical Journal, 138, 1365.
\newblock
  \url{http://stacks.iop.org/1538-3881/138/i=5/a=1365?key=crossref.5d1ef7f7ce2c0bdfaf47b0c994d45b75}

\bibitem[{VanderPlas {et~al.}(2012)VanderPlas, Connolly, Ivezic, \&
  Gray}]{vanderplas_introduction_2012}
VanderPlas, J., Connolly, A.~J., Ivezic, Z., \& Gray, A. 2012, in 2012
  {Conference} on {Intelligent} {Data} {Understanding} (Boulder, CO, USA:
  IEEE), 47--54.
\newblock \url{http://ieeexplore.ieee.org/document/6382200/}

\bibitem[{Wagstaff {et~al.}(2013)Wagstaff, Lanza, Thompson, Dietterich, \&
  Gilmore}]{wagstaff_guiding_2013}
Wagstaff, K.~L., Lanza, N.~L., Thompson, D.~R., Dietterich, T.~G., \& Gilmore,
  M.~S. 2013, in Proceedings of the {Twenty}-{Seventh} {AAAI} {Conference} on
  {Artificial} {Intelligence} {Pages}, 905--911

\bibitem[{Yang \& Li(2015)}]{yang_autoencoder_2015}
Yang, T., \& Li, X. 2015, Monthly Notices of the Royal Astronomical Society,
  452, 158.
\newblock
  \url{https://academic.oup.com/mnras/article-lookup/doi/10.1093/mnras/stv1210}

\bibitem[{Yip {et~al.}(2004{\natexlab{a}})Yip, Connolly, Szalay, Budavári,
  SubbaRao, Frieman, Nichol, Hopkins, York, Okamura, Brinkmann, Csabai, Thakar,
  Fukugita, \& Ivezić}]{yip_distributions_2004}
Yip, C.~W., Connolly, A.~J., Szalay, A.~S., {et~al.} 2004{\natexlab{a}}, The
  Astronomical Journal, 128, 585.
\newblock \url{http://stacks.iop.org/1538-3881/128/i=2/a=585}

\bibitem[{Yip {et~al.}(2004{\natexlab{b}})Yip, Connolly, Vanden~Berk, Ma,
  Frieman, SubbaRao, Szalay, Richards, Hall, Schneider, Hopkins, Trump, \&
  Brinkmann}]{yip_spectral_2004}
Yip, C.~W., Connolly, A.~J., Vanden~Berk, D.~E., {et~al.} 2004{\natexlab{b}},
  The Astronomical Journal, 128, 2603.
\newblock \url{http://stacks.iop.org/1538-3881/128/i=6/a=2603}

\bibitem[{York {et~al.}(2000)York, Adelman, Anderson, Anderson, Annis, Bahcall,
  Bakken, Barkhouser, Bastian, Berman, Boroski, Bracker, Briegel, Briggs,
  Brinkmann, Brunner, Burles, Carey, Carr, Castander, Chen, Colestock,
  Connolly, Crocker, Csabai, Czarapata, Davis, Doi, Dombeck, Eisenstein,
  Ellman, Elms, Evans, Fan, Federwitz, Fiscelli, Friedman, Frieman, Fukugita,
  Gillespie, Gunn, Gurbani, de~Haas, Haldeman, Harris, Hayes, Heckman,
  Hennessy, Hindsley, Holm, Holmgren, Huang, Hull, Husby, Ichikawa, Ichikawa,
  Ivezić, Kent, Kim, Kinney, Klaene, Kleinman, Kleinman, Knapp, Korienek,
  Kron, Kunszt, Lamb, Lee, Leger, Limmongkol, Lindenmeyer, Long, Loomis,
  Loveday, Lucinio, Lupton, MacKinnon, Mannery, Mantsch, Margon, McGehee,
  McKay, Meiksin, Merelli, Monet, Munn, Narayanan, Nash, Neilsen, Neswold,
  Newberg, Nichol, Nicinski, Nonino, Okada, Okamura, Ostriker, Owen, Pauls,
  Peoples, Peterson, Petravick, Pier, Pope, Pordes, Prosapio, Rechenmacher,
  Quinn, Richards, Richmond, Rivetta, Rockosi, Ruthmansdorfer, Sandford,
  Schlegel, Schneider, Sekiguchi, Sergey, Shimasaku, Siegmund, Smee, Smith,
  Snedden, Stone, Stoughton, Strauss, Stubbs, SubbaRao, Szalay, Szapudi,
  Szokoly, Thakar, Tremonti, Tucker, Uomoto, Vanden~Berk, Vogeley, Waddell,
  Wang, Watanabe, Weinberg, Yanny, \& Yasuda}]{york_sloan_2000}
York, D.~G., Adelman, J., Anderson, Jr., J.~E., {et~al.} 2000, The Astronomical
  Journal, 120, 1579.
\newblock \url{http://stacks.iop.org/1538-3881/120/i=3/a=1579}

\bibitem[{Zhao {et~al.}(2019)Zhao, Song, \& Ermon}]{Zhao2019InfoVAEBL}
Zhao, S., Song, J., \& Ermon, S. 2019, in AAAI

\bibitem[{Zhuang {et~al.}(2015)Zhuang, Luo, Jin, Xiong, Luo, \&
  He}]{zhuang_representation_2015}
Zhuang, F., Luo, D., Jin, X., {et~al.} 2015, in 2015 {IEEE} {International}
  {Conference} on {Data} {Mining} (Atlantic City, NJ, USA: IEEE), 1141--1146.
\newblock \url{http://ieeexplore.ieee.org/document/7373449/}

\end{thebibliography}
\bibliographystyle{aasjournal}


\end{document}